\documentclass[journal,draftcls,onecolumn,10pt,twoside]{IEEEtranTCOM} %
\normalsize

\usepackage{cite}

\ifCLASSINFOpdf
  \usepackage[pdftex]{graphicx}
  \usepackage{epstopdf}
  \graphicspath{{figures/}}
\else
\fi

\usepackage[cmex10]{amsmath}
\usepackage{array}

\usepackage{mdwmath}
\usepackage{mdwtab}
\usepackage[caption=false,font=footnotesize]{subfig}
\usepackage{url}

\usepackage{amssymb}
\usepackage{bm}
\usepackage[hidelinks]{hyperref}
\usepackage{microtype}

\makeatletter

\def\underbracexx#1#2{\mathop{\vtop{\m@th\ialign{##\crcr
				$\hfil\displaystyle{#2}\hfil$\crcr
				\noalign{\kern3\p@\nointerlineskip}%
				#1\crcr\noalign{\kern3\p@}}}}\limits}

\def\upbracefilla{$\m@th \setbox\z@\hbox{$\braceld$}%
	\bracelu\leaders\vrule \@height\ht\z@ \@depth\z@\hfill 
	\kern\p@\vrule \@width\p@\kern\p@\vrule \@width\p@\kern\p@\vrule \@width\p@
	$}

\def\upbracefilll{$\m@th \setbox\z@\hbox{$\braceld$}%
	\vrule \@width\p@\kern\p@\vrule \@width\p@\kern\p@\vrule \@width\p@\kern\p@
	\leaders\vrule \@height\ht\z@ \@depth\z@\hfill\braceru$}

\begin{document}
%
\title{Performance Analysis of Multi-User\\Massive MIMO Downlink under\\Channel Non-Reciprocity and Imperfect CSI}
%
%
%

\author{Orod~Raeesi,~\IEEEmembership{Student~Member,~IEEE,}
        Ahmet~Gokceoglu,~\IEEEmembership{Member,~IEEE,}
        Yaning~Zou,~\IEEEmembership{Member,~IEEE,}
        Emil~Bj\"{o}rnson,~\IEEEmembership{Senior Member,~IEEE,}
        and~Mikko~Valkama,~\IEEEmembership{Senior Member,~IEEE}
\thanks{O. Raeesi, A. Gokceoglu, and M. Valkama are  with the Department of Electronics and Communications Engineering, Tampere University of Technology, Tampere 33720, Finland (e-mail: orod.raeesi@tut.fi; ahmet.gokceoglu@tut.fi; mikko.e.valkama@tut.fi).}
\thanks{Y. Zou is with Technische Universit\"{a}t Dresden, Vodafone Chair Mobile Communications Systems, Dresden, Germany (e-mail: yaning.zou@ifn.et.tu-dresden.de).}
\thanks{E. Bj\"{o}rnson is with the Department of Electrical Engineering (ISY), Link̈\"{o}ping University, SE-581 83 Link\"{o}ping, Sweden (e-mail:emil.bjornson@liu.se).}
\thanks{This work was supported by the Finnish Funding Agency for Technology and Innovation (Tekes) under the project ``$5$th Evolution Take of Wireless Communication Networks (TAKE-$5$)'', by the Academy of Finland under the projects $284694$ and $288670$ and TUT Graduate School.}
}
%
%

\markboth{\tiny Raeesi \MakeLowercase{\textit{et al.}}: Performance Analysis of Multi-User Massive MIMO Downlink under Channel Non-Reciprocity and Imperfect CSI}%
{\tiny Raeesi \MakeLowercase{\textit{et al.}}: Performance Analysis of Multi-User Massive MIMO Downlink under Channel Non-Reciprocity and Imperfect CSI}
%



\date{}

\maketitle

\begin{abstract}
This paper analyzes the performance of linearly precoded time division duplex based multi-user massive MIMO downlink system under joint impacts of channel non-reciprocity (NRC) and imperfect channel state information (CSI). We consider a {generic and realistic} NRC model that accounts for transceiver frequency-response {as well as mutual coupling mismatches at both user equipment (UE) and base station (BS) sides}. The analysis covers two most prominent forms of linear precoding schemes, namely, zero-forcing (ZF) and maximum-ratio transmission (MRT), and assumes that only the statistical {properties of the beamformed channel} are used at the UE side to decode the received signal. {Under the approximation of i.i.d. Gaussian channels, closed-form} analytical expressions are derived for the effective signal to interference and noise ratios (SINRs) and the corresponding capacity lower bounds. The expressions show that, in moderate to high SNR, the additional interference caused by {imperfect NRC calibration can degrade} the performance of both precoders significantly. Moreover, ZF is shown to be more sensitive to NRC than MRT. Numerical evaluations with practical NRC levels indicate that this performance loss in the spectral efficiency can be as high as $42$\% for ZF, whereas it is typically less than $13$\% for MRT. { It is also shown that due to the NRC, the asymptotic large-antenna performance of both precoders saturate to an identical finite level.} The derived analytical expressions provide useful tools and valuable technical insight, e.g., into calculating the NRC calibration requirements in BSs and UEs for any given specific performance targets in terms of effective SINR or the system capacity bound.
\end{abstract}

\begin{IEEEkeywords}
Capacity, channel reciprocity, frequency-response mismatch, inter-user interference, linear precoding, multi-user massive MIMO, mutual coupling, SINR.
\end{IEEEkeywords}

\IEEEpeerreviewmaketitle

\section{Introduction}
\IEEEPARstart{M}{assive} multiple-input multiple-output (MIMO) systems are envisioned to be one key enabling technology for the next generation cellular networks, known as $5$G \cite{5G_2,5G}. In massive MIMO systems, a base station (BS) uses an array with a large number of antennas $N$ to serve $K$ user equipments (UEs) simultaneously on the same time-frequency resource, where typically $N \gg K$ \cite{5G,massMimo0,massMIMO01,Marzetta0}. Large-scale system analysis shows that linear precoding techniques, e.g., zero-forcing (ZF) and maximum ratio transmission (MRT) are asymptotically optimal with increasing $N$, while very high spectral-efficiencies can already be achieved with $N$ being in the order of several tens or hundreds \cite{massMIMOeffic,Marzetta2,massMIMOchannel,Marzetta0}.

The key requirement for employing the above precoding schemes is to have accurate channel state information (CSI) at the BS for efficient multi-user spatial precoding. In conventional frequency-division duplex (FDD) based MIMO systems, where the number of BS antennas is relatively low, UEs commonly estimate downlink (DL) channels based on the received DL training signals transmitted by the BS, and feed the estimated DL channels back to the BS \cite{FDD}. The number of DL pilots required for estimating the channels is proportional to the number of antennas in the BS which complicates the adoption of such DL channel estimation and reporting methods in massive MIMO systems. As an alternative approach, massive MIMO systems are typically assumed to employ time-division duplex (TDD), and thus estimate the DL channel based on uplink (UL) pilots, relying on the reciprocity of the physical DL and UL channels within channel coherence interval \cite{TDDprefer}. Thereby, the required amount of resources in such a TDD based approach is only proportional to the number of served UEs which is typically much smaller than the number of BS antennas, i.e., $K \ll N$ \cite{TDDprefer,Marzetta0}.

The channel reciprocity in TDD systems holds only for the physical propagation channels. However, when the effective baseband-to-baseband transmission channels between the BS and UEs are considered, incorporating also the impacts of the involved transceiver circuits {and antenna systems}, the reciprocity does not hold anymore due to the mismatches in transmit and receive mode characteristics of the transceivers {and antenna systems} \cite{nonRecip,nonRecip3,JianProblem,Kouassi}. More specifically, such mismatch characteristics include the unavoidable differences between the frequency-responses (FRs) of transmitter and receiver chains of any individual transceiver, as well as the mutual coupling effects between the antenna elements in multi-antenna devices \cite{Petermann,ourVTC14Analysis,MC}. 
The impacts of such transceiver hardware {and antenna system} induced non-reciprocity, also commonly referred to as channel non-reciprocity (NRC), have been studied for massive MIMO systems to a certain extent in \cite{ourGlobecom14,idealnonRecip1,idealnonRecip2,idealnonRecip3,nonRecipinMassMIMO}. To this end, {\cite{ourGlobecom14,idealnonRecip1,idealnonRecip2,idealnonRecip3,nonRecipinMassMIMO}} study the system performance degradation in terms of signal to interference and noise ratios (SINRs) and the corresponding achievable rates due to NRC, while assuming otherwise ideal system with perfect CSI. Furthermore, the system models in \cite{idealnonRecip2,idealnonRecip3,nonRecipinMassMIMO} consider only FR mismatch {and thus ignore} the NRC induced by possible mutual coupling mismatches, reported, e.g., in \cite{Petermann,ourVTC14Analysis,MC,idealnonRecip1} to be one important practical source of non-reciprocity. Furthermore, only the BS side NRC is considered in \cite{idealnonRecip1,idealnonRecip2}.

In this paper, we analyze the SINR and achievable sum-rate of linearly precoded TDD multi-user massive MIMO DL transmission systems under the joint impacts of imperfect CSI and NRC. We consider a generic and realistic NRC model which takes into account {both the FR and mutual coupling mismatches at the UEs and the BS}. The analysis is carried out for the two most widely-adopted forms of linear precoding, namely, ZF and MRT. As in \cite{Marzetta2,nonRecipinMassMIMO,ZF,Marzetta3}, we also assume that UEs rely only on statistical DL CSI to decode the received signals{, and thus more sophisticated precoding schemes, e.g., block diagonalization-based precoding, requiring instantaneous {demodulation CSI at UE receivers} are excluded}. Based on the developed signal and system models, closed-form expressions are derived for the effective SINRs and the corresponding capacity lower bounds. To highlight the substantial differences between this work and the existing literature on performance analysis of NRC impaired massive MIMO systems, we summarize the novel contributions of this manuscript as follows:
{\begin{enumerate}
\item In contrast to the simplified NRC {{models in \cite{idealnonRecip2,idealnonRecip3,nonRecipinMassMIMO} which consider}} only FR mismatches, a more practical and generic NRC model is considered in this work which incorporates both FR and mutual coupling mismatches in both BS and UE sides.
\item {{In contrast to the existing literature, the analysis in this work does not impose any restrictions on the structure of NRC matrices and the involved NRC variables, in terms of their statistical distributions or mutual correlation. Therefore, in addition to covering the systems without explicit NRC calibration, the provided analytical results can also be used in connection with  residual non-reciprocity after any given NRC calibration method, e.g., \cite{shepard_argos:_2012,recipTemp,Rogalin_rel_calib}.}}
\item In contrast to \cite{idealnonRecip1,idealnonRecip3}, a performance comparison between ZF and MRT precoding schemes is also carried out which shows the relative sensitivity of these precoders to different NRC levels, with and without UL channel estimation errors, in both non-asymptotic and asymptotic cases.
\item In contrast to {{\cite{ourGlobecom14,idealnonRecip1,idealnonRecip2,idealnonRecip3,nonRecipinMassMIMO}}} which consider NRC alone, in this work we consider the joint impacts of co-existing NRC and UL channel estimation errors (called imperfect CSI).
\item In contrast to \cite{ourGlobecom14,idealnonRecip1,idealnonRecip2,idealnonRecip3,nonRecipinMassMIMO}, the derived analytical expressions decompose the total received interference into two parts, namely, interference power due to imperfect CSI, without NRC, and the interference term due to NRC (see expression \eqref{ZF SINR} for ZF, and \eqref{MRT SINR} for MRT). With this decomposition, it is straightforward to quantify the specific performance degradation due to NRC with respect to the ideal reciprocal case, and also to draw technical insight and establish design criteria for both UL pilot signaling and reciprocity calibration.
\end{enumerate}}
{{{In general, given the specific}} performance targets, such as effective SINRs and/or capacity lower bound, the derived analytical expressions reported in this manuscript can be directly used in designing and dimensioning the system, e.g., choosing the appropriate precoder based on the performance-complexity trade-off, deciding on the number of active antenna elements, and/or extracting the needed accuracy of NRC calibration schemes{, as well as understanding the trade-offs between UL pilot based channel estimation accuracy, NRC calibration accuracy and the achievable system performance.}}

The rest of the paper is organized as follows. Section \ref{sec:sysModel} describes the fundamental multi-user massive MIMO system model under transceiver {and antenna system} non-reciprocity and imperfect CSI. Then, in Section \ref{sec:analysis}, analytical expressions are derived for the effective DL SINR and capacity lower bound under ZF and MRT precoding schemes. In Section \ref{sec:implications}, the asymptotic SINR and achievable rate expressions are derived for ZF and MRT precoding schemes, and also an analytic performance comparison is pursued in both asymptotic and non-asymptotic cases. In Section \ref{sec:results}, extensive numerical results are provided to evaluate and verify the derived analytical expressions and illustrate the impact of various non-reciprocity sources and parameters on the system performance. Finally, conclusions are drawn in Section \ref{sec:conclusions}. Selected details regarding the derivations of the reported analytic expressions are provided in an Appendix.

\textit{Notations:} Throughout this paper, matrices (vectors) are denoted with upper{-case} (lower{-case}) bold characters, e.g., $\mathbf{V}$ ($\mathbf{v}$). The superscripts $\left(.\right)^\mathrm{T}$, $\left(.\right)^*$, and $\left(.\right)^\mathrm{H}$ stand for transpose, conjugate, and conjugate-transpose, respectively. Expectation operator is shown by $\mathbb{E}{\left[.\right]}$, $\mathrm{Tr}\left(.\right)$ represents the trace operator, {$\mathrm{Sum}\left(.\right)$ yields the element-wise sum of the argument matrix, }while $\mathrm{Var}\left(.\right)$ and {$\mathrm{Cov}\left(.\right)$} refer to the variance and {covariance} operators, respectively. $\mathbf{I}_n$ and $\mathbf{0}_n$ denote $n\times n$ identity and all-zero matrices, respectively. The element in $i$-th row and $j$-th column of matrix $\mathbf{V}$ is represented by $v_{ij}$. A diagonal matrix with elements $\left(v_{1}, \cdots, v_{n}\right)$ is shown by $\mathrm{diag}\left(v_{1}, \cdots, v_{n}\right)$, {corresponding block-diagonal matrix is denoted by $\mathrm{blkdiag}\left(\mathbf{A}_{1}, \cdots, \mathbf{A}_{k}\right)$,} and $\mathcal{CN}\left(0,\sigma^2 \right)$ represents a circularly symmetric zero-mean complex Gaussian distribution with variance $\sigma^2$.

\section{System Model} \label{sec:sysModel}
We consider precoded downlink data transmission in a TDD based multi-user massive MIMO system, where a BS with $N$ antennas serves {$K$ UEs} simultaneously on the same time-frequency resource. {The number of antennas in $k$-th UE is denoted by $M_k$ and $\sum_{i = 1}^K M_k = M_{tot}$, where $N \gg M_{tot}$. For notational convenience, we assume that the total set of $M_{tot}$ antennas at the UE side is logically indexed such that the first $M_1$ antennas belong to UE $1$, the next $M_2$ antennas belong to UE $2$, and so forth}. {We also assume that all antenna elements in the considered system are omni-directional, for simplicity.} We further assume that the spatial transmit signal vector is generated using linear precoding techniques, e.g., ZF or MRT. All system models are written for an arbitrary subcarrier of the underlying orthogonal frequency division multiplexing/multiple access (OFDM/OFDMA) waveform, that is, before IFFT and after FFT on the TX and RX sides, respectively, without explicitly showing the subcarrier index. It is further assumed that the cyclic prefix (CP) length is larger than the channel delay spread.

\subsection{Uplink Training, Downlink Transmission and Effective Channels}
The {DL} linear precoder is designed based on the CSI acquired from UL pilots. The fundamental {multi-user} signal models for the UL pilot and DL data transmission phases can be expressed as \cite{massMIMOeffic,Marzetta1}
\begin{equation} \label{eq:sys Model}
\begin{aligned}
&\mathrm{UL}: \mathbf{Y}^p = \sqrt{{\tau_u}\rho_u}\mathbf{G}\mathbf{X}^p + \mathbf{N}^p\\
&\mathrm{DL}: \mathbf{r} = \sqrt{\rho_d}\mathbf{H}\mathbf{x} + \mathbf{n},
\end{aligned}
\end{equation}
where $\mathbf{G} \in \mathbb{C}^{N \times {M_{tot}}}$ and $\mathbf{H} \in \mathbb{C}^{{M_{tot}} \times N}$ are the effective UL and DL channel matrices, respectively, that are explicitly defined in the next paragraph. Regarding the UL {pilot} signal model, $\rho_u$ is the transmitted signal to noise ratio (SNR) of the UL pilots, ${\mathbf{Y}^p = \left[\mathbf{y}^{p}_{1}, \cdots, \mathbf{y}^{p}_{N}\right]^\mathrm{T}}$ is the received signal matrix at the BS receiver, stacking the received UL pilots over $\tau_u$ symbol durations, where $\mathbf{y}^{p}_{n} \in \mathbb{C}^{\tau_u \times 1}$ contains the received UL pilots at $n$-th BS antenna, and $\mathbf{N}^p = {\left[\mathbf{n}^{p}_{1}, \cdots, \mathbf{n}^{p}_{N}\right]^\mathrm{T}}$ is the additive receiver noise matrix at the BS with i.i.d. $\mathcal{CN}\left(0,1\right)$ elements, where $\mathbf{n}^{p}_{n} \in \mathbb{C}^{\tau_u \times 1}$ is the additive receiver noise {sequence} at $n$-th BS antenna. The matrix stacking all the transmitted UL pilots at all the {antennas in the UE side} is shown by $\mathbf{X}^p = \left[\mathbf{x}^{p}_{1}, \cdots, \mathbf{x}^{p}_{{M_{tot}}}\right]^\mathrm{T}$, where ${\mathbf{x}^{p}_{m}} \in \mathbb{C}^{\tau_u \times 1}$ is the UL temporal pilot vector transmitted from {$m$}-th {antenna in the UE side}. Then, for the DL, $\mathbf{r} \in \mathbb{C}^{{M_{tot}} \times 1}$ denotes the received {multi-user} DL signal vector corresponding to all {$M_{tot}$ antennas at the UE side}, $\rho_d$ is the transmitted SNR of DL channel, and $\mathbf{n} \in \mathbb{C}^{{M_{tot}} \times 1}$ is the normalized additive receiver noise vector at UE side with i.i.d. $\mathcal{CN}\left(0,1\right)$ elements. The precoded spatial transmit signal vector in the BS is shown by $\mathbf{x} = \left[x_1, \cdots, x_N\right]^\mathrm{T}$, where $x_n$ is the precoded sample transmitted from $n$-th antenna in the BS.

\begin{figure*}[!t]
\centering
\subfloat[]{\includegraphics[height=0.26\textheight]{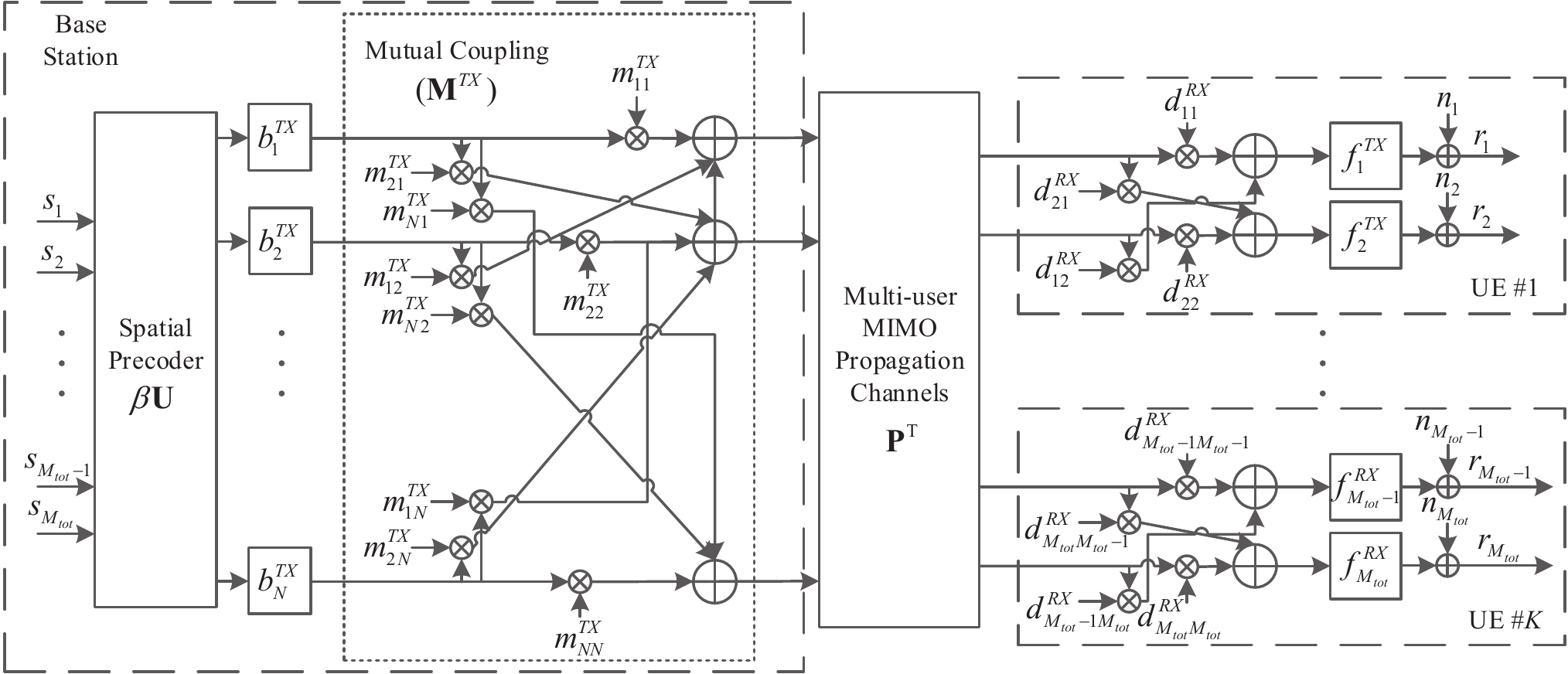}
\label{subfig:DL}}
\hfil
\subfloat[]{\includegraphics[height=0.26\textheight]{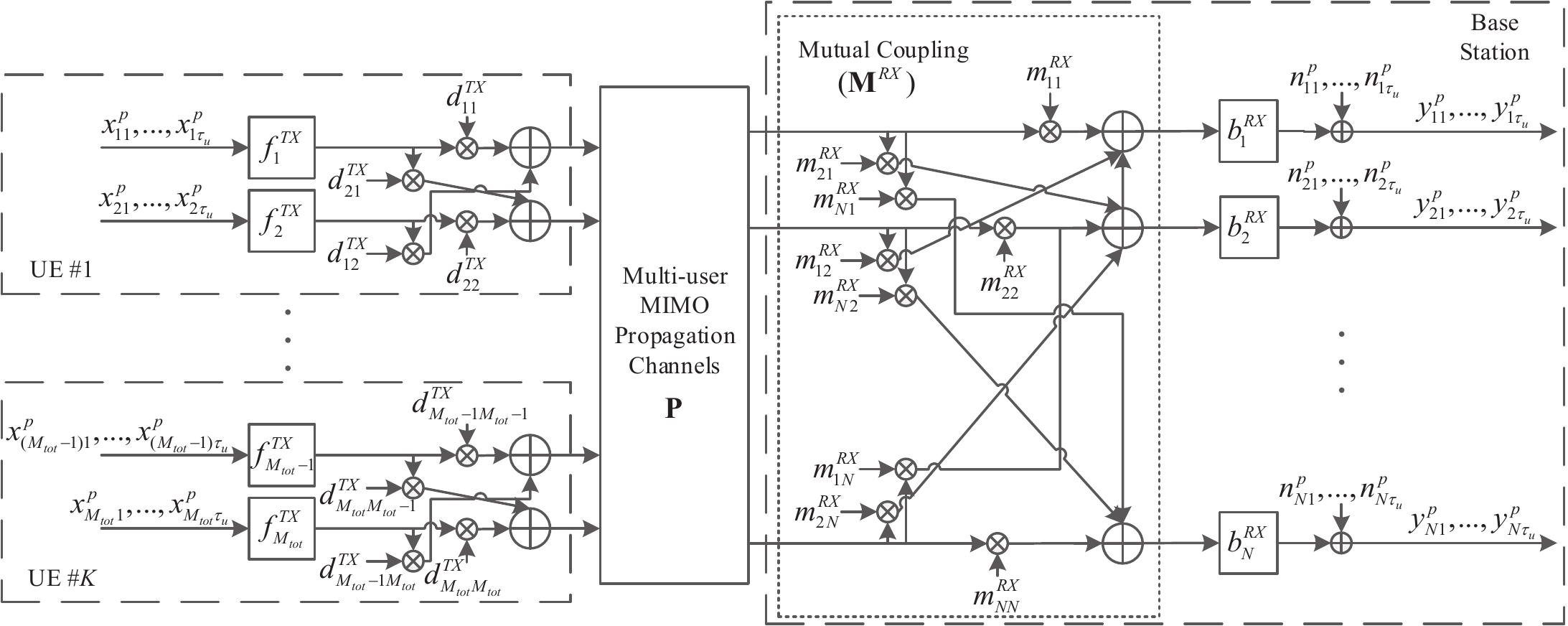}
\label{subfig:UL}}
\caption{Principal illustration of \protect\subref{subfig:DL} DL and \protect\subref{subfig:UL} UL transmissions and receptions including physical propagation channels, transceiver frequency responses and antenna mutual coupling in the devices {in an example case of dual-antenna UEs.}}
\label{fig:eff chan}
\end{figure*}
As illustrated in \figurename{~\ref{fig:eff chan}}, the effective DL and UL channels are generally cascades of transceiver frequency-responses and antenna mutual coupling at BS side, physical propagation channels, and {transceiver frequency-responses and antenna mutual coupling} at UE side. Thus, the effective DL channel $\mathbf{H}$ and the effective UL channel $\mathbf{G}$ can be written explicitly as \cite{Petermann,ourVTC14Analysis}
\begin{equation} \label{eq:UL and DL}
\begin{gathered} 
\mathbf{H} = \mathbf{F}^{RX} {\mathbf{D}^{RX}} \mathbf{P}^\mathrm{T} \mathbf{M}^{TX} \mathbf{B}^{TX} \\
\mathbf{G} = \mathbf{B}^{RX} \mathbf{M}^{RX} \mathbf{P} {\mathbf{D}^{TX}} \mathbf{F}^{TX},
\end{gathered}
\end{equation}
where $\mathbf{F} = \mathrm{diag}\left(f_1, \cdots, f{_{M_{tot}}}\right)$ is the {total} FR matrix of the UEs, {$\mathbf{D}  = \mathrm{blkdiag}\left(\mathbf{D}_1, \cdots, \mathbf{D}_K\right) \in \mathbb{C}^{M_{tot} \times M_{tot}}$ is a block-diagonal matrix representing the antenna mutual coupling matrix at UE side}, $\mathbf{B} = \mathrm{diag}\left(b_1, \cdots, b_N\right)$ is the FR matrix of the BS, $\mathbf{M} \in \mathbb{C}^{N \times N}$ is the antenna mutual coupling matrix of the BS, and $\mathbf{P} \in \mathbb{C}^{N \times {M_{tot}}}$ is the reciprocal physical channel, while the superscripts $TX$ and $RX$ specify the transmit and receive modes, respectively. Notice that while the overall UE side antenna mutual coupling matrices, $\mathbf{D}^{TX}$ and $\mathbf{D}^{RX}$, are assumed to be block-diagonal, because of clear physical separation of the different UE devices, the element matrices $\mathbf{D}^{TX}_k$ and $\mathbf{D}^{RX}_k$ are generally full matrices of size $M_{k} \times M_{k}$.

\subsection{Channel Non-Reciprocity Problem}
As outlined above, in TDD networks the BS obtains DL CSI based on the estimated UL channel, since DL and UL channels share the same spectrum and are assumed to be reciprocal within each channel coherence interval. The reciprocal nature applies, however, only to the physical propagation channels shown in \figurename{~\ref{fig:eff chan}}. In addition to the physical channels, the effective channels also include the responses of electronics components used in the transmitting and receiving devices which results into the effective DL and UL channels expressed in \eqref{eq:UL and DL}.

Based on \eqref{eq:UL and DL}, the relation between the effective DL and UL channels can now be expressed as
\begin{equation} \label{eq:non-recip}
\mathbf{H} = \mathbf{A} \mathbf{G}^\mathrm{T} \mathbf{C},
\end{equation}
where the matrices $\mathbf{A}$ and $\mathbf{C}$ are
\begin{equation} \label{eq:non-recip2}
\begin{gathered} 
\mathbf{A} = \mathbf{F}^{RX} {\mathbf{D}^{RX} \left({\mathbf{D}^{TX}}\right)^{-\mathrm{T}}} \left(\mathbf{F}^{TX}\right)^{-1} \\
\mathbf{C} = \left({\mathbf{B}^{RX}}\right)^{-1} \left({\mathbf{M}^{RX}}\right)^{-{\mathrm{T}}} \mathbf{M}^{TX}\mathbf{B}^{TX}.
\end{gathered}
\end{equation}
In \eqref{eq:non-recip} and \eqref{eq:non-recip2}, the matrices $\mathbf{A} \in \mathbb{C}^{{M_{tot}} \times {M_{tot}}}$ and $\mathbf{C} \in \mathbb{C}^{N \times N}$ are incorporating the effects of transceivers {and antenna systems} on the non-reciprocity in UEs and BS, respectively. The matrix $\mathbf{A}$ is block-diagonal and can in general be written as $\mathbf{A} = \mathbf{I}{_{M_{tot}}} + \mathbf{A}^\prime$ where $\mathbf{A}^\prime$ can be expressed as $\mathbf{A}^\prime = \mathrm{blkdiag}\left(\mathbf{A}^\prime_1, \cdots, \mathbf{A}^\prime_K\right)$, while the full matrix $\mathbf{A}^\prime_k \in \mathbb{C}^{M_k \times M_k}$ represents the NRC in the $k$-th UE. On the other hand, $\mathbf{C}$ which represents the overall BS transceiver {and antenna system} non-reciprocity, including mutual coupling mismatch, is generally {an $N \times N$} full matrix and can be decomposed as ${\mathbf{C} = \mathbf{I}_{N} + \mathbf{C}^\prime}$.

In general, the channel non-reciprocity values vary very slowly in time with respect to the variations in the propagation channel \cite{recipTemp} and hence $\mathbf{A}$ and $\mathbf{C}$ can be assumed to remain constant over many channel coherence intervals. Furthermore, it can easily be deduced that the effective DL and UL channels are reciprocal if and only if the mismatch matrices satisfy $\mathbf{A}^\prime = \mathbf{0}{_{M_{tot}}}$ and $\mathbf{C}^\prime = \mathbf{0}_N$.

{For the purpose of the upcoming analysis, we next define and assume the following. First, we write $\mathbf{A}^\prime_k$ as $\mathbf{A}^\prime_k = \left[{\mathbf{a}^\prime}^k_1, \cdots, {\mathbf{a}^\prime}^k_{M_k}\right]^\mathrm{T}$ and by dropping the UE index $k$ for notational simplicity, we define $\mathbf{R}_{\mathbf{a}^\prime_{m}} = \mathrm{Cov}\left(\mathbf{a}^\prime_{m}\right)$ for the $m$-th antenna at the UE side ranging from $1$ to $M_{tot}$. In matrix ${\mathbf{A}^\prime}$, the elements are assumed to be zero-mean and the power of $a^\prime_{mi}$ is denoted by ${\sigma_{{a}^\prime_{mi}}^2 = \mathbb{E}{\left[\left|a_{mi}^\prime\right|^2\right]}}$. Similarly at the BS side, $\mathbf{C}^\prime$ is also assumed to have zero-mean elements. Then, we stack all the diagonal and non-diagonal elements of $\mathbf{C}^\prime$ in $\mathbf{c^\prime_d} = \left[c^\prime_{11}, c^\prime_{22}, \cdots, c^\prime_{NN}\right]$ and $\mathbf{c^\prime_{od}} = \left[c^\prime_{12}, c^\prime_{13}, \cdots, c^\prime_{NN-1}\right]$, respectively, and define ${\mathbf{R}_{\mathrm{c}^\prime_d} = \mathrm{Cov}\left(\mathbf{c^\prime_d}\right)}$ and $\mathbf{R}_{\mathrm{c}^\prime_{od}} = \mathrm{Cov}\left(\mathbf{c^\prime_{od}}\right)$. Then, as explicitly shown in the appendix, the final closed-form analysis results depend only on these NRC covariances but not, e.g., on the exact distributions of the NRC variables. {In all the forth-coming analysis and derivations, we adopt the simplifying assumption or approximation that the elements of the effective UL channel $\mathbf{G}$ are unit-variance i.i.d. Gaussians. While the exact distribution and correlation characteristics of real-world effective UL channel entries depend, among others, on the exact antenna array configuration and angular spread of the propagation environment, we adopt such simplifying approximation since the closed-form rate expressions that one can deduce by using such model have been shown to match very accurately with practical massive MIMO measurements \cite{massMIMOtut}. This is a result of the channel hardening and favorable propagation phenomena, which makes the performance less dependent on the actual channel distribution. Hence, we use i.i.d. Rayleigh fading in this work to study the channel non-reciprocity aspects in a clean and rigorous manner.}}

\subsection{Channel Estimation}
To facilitate the channel estimation at the BS, the UEs simultaneously transmit mutually orthogonal UL pilot sequences of length $\tau_u$ such that ${{\mathbf{X}^p \left({\mathbf{X}^p}\right)^\mathrm{H}}} = \mathbf{I}{_{M_{tot}}}$ with $\tau_u \geq {M_{tot}}$.
To estimate the UL channels, the BS multiplies $\mathbf{Y}^p$ in \eqref{eq:sys Model} by $\left({\mathbf{X}^p}\right)^\mathrm{H}$, which yields \cite{Marzetta2}
\begin{equation}
\mathbf{Y} = \mathbf{Y}^p \left({\mathbf{X}^p}\right)^\mathrm{H} = \sqrt{\tau_u \rho_u} \mathbf{G} + \mathbf{Q},
\end{equation}
where $\mathbf{Q} \in \mathbb{C}^{N \times {M_{tot}}}$ is the processed noise matrix with i.i.d. $\mathcal{CN}\left(0,1\right)$ elements. Using minimum mean-square error (MMSE) channel estimator, the estimated effective UL channel $\hat{\mathbf{G}} \in \mathbb{C}^{N \times {M_{tot}}}$ can be shown to read \cite{Marzetta2,Marzetta1}
\begin{equation} \label{eq:DLchanEst}
\hat{\mathbf{G}} {\ = \frac{\sqrt{\tau_u \rho_u}}{\tau_u \rho_u + 1} \mathbf{Y}} = \frac{\tau_u \rho_u}{\tau_u \rho_u + 1}{\mathbf{G}} + \frac{\sqrt{\tau_u \rho_u}}{\tau_u \rho_u + 1}\mathbf{Q},
\end{equation}
while the corresponding effective DL channel estimate, called $\hat{\mathbf{H}}$, that is utilized by the NRC-unaware BS for downlink precoding is obtained by $\hat{\mathbf{H}} = \hat{\mathbf{G}}^\mathrm{T}$. Based on \eqref{eq:DLchanEst} and the orthogonality principle of MMSE estimators, the effective UL channel matrix $\mathbf{G}$ can also be decomposed as \cite{Marzetta2,Marzetta1}
\begin{equation} \label{eq:ULchanEst}
{\mathbf{G}} = \hat{\mathbf{G}} + \bm{\mathcal{E}}^\mathrm{T} = \hat{\mathbf{H}}^\mathrm{T} + \bm{\mathcal{E}}^\mathrm{T},
\end{equation}
where $\bm{\mathcal{E}} = \left[{\bm{\varepsilon}}_1,...,{\bm{\varepsilon}}{_{M_{tot}}}\right]^\mathrm{T} \in \mathbb{C}^{{{M_{tot}}} \times N}$ accounts for the UL channel estimation errors and has i.i.d. $\mathcal{CN}\left(0,\frac{1}{\tau_u \rho_u + 1}\right)$ elements. The estimated effective DL channel $\hat{\mathbf{H}}$ has i.i.d. $\mathcal{CN}\left(0,\frac{\tau_u \rho_u}{\tau_u \rho_u + 1}\right)$ elements and is independent of $\bm{\mathcal{E}}$. {The considered pilot signaling and UL channel estimation method is the most common form of UL CSI acquisition for massive MIMO systems in the existing literature \cite{Marzetta2,massMIMOeffic,massMimo0}.} {Alternative partial CSI acquisition based approaches, such as \cite{partial_CSI}, are also important but are outside the scope of this paper.}

Incorporating \eqref{eq:ULchanEst} into \eqref{eq:non-recip}, we finally obtain the relation between the estimated and true effective DL channels as
\begin{equation} \label{eq:H and H hat}
\mathbf{H} = \mathbf{A} \mathbf{G}^\mathrm{T} \mathbf{C} = \mathbf{A} \left(\hat{\mathbf{H}} + \bm{\mathcal{E}}\right) \mathbf{C},
\end{equation}
which summarizes the joint effects of two co-existing non-ideality sources, namely, UL channel estimation error and the channel non-reciprocity, on the effective DL channel estimation.

\section{Performance Analysis under NRC and Imperfect CSI} \label{sec:analysis}
In this section, we characterize the impacts of {coexisting} NRC and imperfect CSI on the performance of linearly precoded multi-user massive MIMO DL transmission. In this respect, we will derive analytical expressions for the received SINR and achievable rates for both ZF and MRT precoding.

\subsection{Downlink Received Signal Model and SINR}
We first express the linearly precoded DL transmit vector $\mathbf{x} \in \mathbb{C}^{N \times 1}$ as
\begin{equation} \label{eq:precoded}
\mathbf{x} = \beta\mathbf{U}\mathbf{s},
\end{equation}
where $\mathbf{U} = \left[\mathbf{u}_1,...,\mathbf{u}{_{M_{tot}}}\right] \in \mathbb{C}^{N \times {{M_{tot}}}}$ is the precoder matrix. The normalized multi-user data vector including one stream per UE antenna is denoted by ${\mathbf{s} = \left[s_1,...,s{_{M_{tot}}}\right]^\mathrm{T} \in \mathbb{C}^{{{M_{tot}}} \times 1}}$, where {$\mathbb{E}\left[\mathbf{s}\mathbf{s}^\mathrm{H}\right] = \mathbf{I}_{M_{tot}}$}. The transmit {sum-power} normalization is achieved through $\beta$ which constrains the total BS transmit {sum-power} to $1$, i.e., $\mathbb{E}[\mathbf{x}^\mathrm{H}\mathbf{x}] = 1$. In order to satisfy {this condition}, $\beta$ is chosen as \cite{Marzetta1}
\begin{equation} \label{eq:beta}
\beta = \left(\sqrt{\mathbb{E}{\left[{\mathrm{Tr}\left({\mathbf{U}}^\mathrm{H}{\mathbf{U}}\right)}\right]}}\right)^{-1} .
\end{equation}

Substituting \eqref{eq:precoded} in {\eqref{eq:sys Model}}, the received DL {multi-user signal vector} corresponding to all {$M_{tot}$ antennas in the UE side} reads
\begin{equation} \label{eq:received0}
\mathbf{r} = \beta\sqrt{\rho_d}\mathbf{H}\mathbf{U}\mathbf{s} + \mathbf{n}.
\end{equation}
We express the effective DL channel matrix as $\mathbf{H} = \left[\mathbf{h}_1,...,\mathbf{h}{_{M_{tot}}}\right]^\mathrm{T}$, where ${\mathbf{h}_m^\mathrm{T}}$ is the effective DL channel from the BS to the {$m$}-th {antenna at the UE side}. Then, based on \eqref{eq:H and H hat} and \eqref{eq:received0}, the received DL signal at the {$m$}-th UE antenna, which is assumed to belong to UE $k$, can be expressed as
\begin{equation} \label{eq:received1}
\begin{aligned}
{r_m} &= \sqrt{\rho_d}{\beta}\mathbf{h}_m^\mathrm{T}{{\mathbf{u}}}_{m}s_m + \sqrt{\rho_d}{\beta}\sum\limits_{{i = 1, i \ne m}}^{M_{tot}} {\mathbf{h}}_m^\mathrm{T}{{\mathbf{u}}}_{i}s_i  + n_m \\
&= \sqrt{\rho_d}{\beta}\sum\limits_{l \in \mathrm{UE}_k} a_{ml} \left(\hat{\mathbf{h}}_l^\mathrm{T} + {\bm{\varepsilon}}_l^\mathrm{T}\right) \mathbf{C}{{\mathbf{u}}}_{m}s_m + \sqrt{\rho_d} {\beta} \sum\limits_{\substack{i = 1,i \ne m}}^{M_{tot}} \sum\limits_{l \in \mathrm{UE}_k} a_{ml} \left(\hat{\mathbf{h}}_l^\mathrm{T} + {\bm{\varepsilon}}_l^\mathrm{T}\right) \mathbf{C}{{\mathbf{u}}}_{i}s_i  + n_m,
\end{aligned}
\end{equation}
where $\mathrm{UE}_k$ refers to the set of logical antenna indices belonging to UE $k$.

Similar to \cite{Marzetta2,nonRecipinMassMIMO,ZF,Marzetta3}, we assume that the UEs rely only on the statistical properties of the {beamformed channel} to decode the received DL signal, i.e., the $k$-th UE uses only ${\beta}\mathbb{E}\left[{\mathbf{h}_m^\mathrm{T}}\mathbf{u}{_{m}}\right]$ as the DL complex gain in {detecting $s_m$}. Therefore, the received signal {in \eqref{eq:received1}} can be decomposed as {
\begin{equation} \label{eq:received2}
{r_m} = \underbrace{\sqrt{\rho_d}{\beta}\mathbb{E}\left[\mathbf{h}_m^\mathrm{T}{{\mathbf{u}}}_{m}\right]s_m}_{\mathrm{useful \ signal}} + z_m^\mathrm{SI} + z_m^\mathrm{ISI} + n_m,
\end{equation}}
where ${z_m^\mathrm{SI}}$ and ${z_m^\mathrm{ISI}}$ are the self-interference (SI) and {inter-stream interference (ISI)}, respectively, which can be explicitly expressed as
\begin{equation}\label{eq:SI}
\begin{gathered}
z_{m}^\mathrm{SI} = \sqrt{\rho_d}{\beta}\sum\limits_{l \in \mathrm{UE}_k} a_{ml} \left(\hat{\mathbf{h}}_l^\mathrm{T} + {\bm{\varepsilon}}_l^\mathrm{T}\right) \mathbf{C}{{\mathbf{u}}}_{m}s_m - \sqrt{\rho_d}{\beta}\mathbb{E}\left[\mathbf{h}_m^\mathrm{T}{{\mathbf{u}}}_{m}\right]s_m\\
z_{m}^\mathrm{ISI} = \sqrt{\rho_d}{\beta}\sum\limits_{\substack{i = 1,i \ne m}}^{M_{tot}} \sum\limits_{l \in \mathrm{UE}_k} a_{ml} \left(\hat{\mathbf{h}}_l^\mathrm{T} + {\bm{\varepsilon}}_l^\mathrm{T}\right) \mathbf{C}{{\mathbf{u}}}_{i}s_i.
\end{gathered}
\end{equation}
Note that, in this definition, the ISI consists of both inter-stream interference from other streams targeted to the same UE and of inter-user interference (IUI) due to the streams of other UEs.

Based on \eqref{eq:received2}, the effective SINR {at the $m$-th antenna in the UE side} can be written as{
\begin{equation} \label{SINR}
\mathrm{SINR}_m = \frac{\mathrm{Var}\left(\mathrm{\sqrt{\rho_d}{\beta}\mathbb{E}\left[\mathbf{h}_m^\mathrm{T}{{\mathbf{u}}}_{m}\right]s_m}\right)}{\mathrm{Var}\left(z_m^\mathrm{SI}\right)+\mathrm{Var}\left(z_m^{{\mathrm{ISI}}}\right)+1},
\end{equation}}
where in defining \eqref{SINR} we used the fact that {$z_m^\mathrm{SI}$ and $z_m^\mathrm{ISI}$} are uncorrelated.

In deriving capacity lower bounds, we follow the same approach as in \cite{Marzetta2,Marzetta4}. The total noise/interference term is uncorrelated with the useful signal whose entropy is upper-bounded with the entropy of Gaussian noise with equal variance \cite{lowerBound}. Hence, a lower-bound on the achievable sum-capacity can be expressed as
\begin{equation}\label{eq:rate}
R = {\sum\limits_{m = 1}^{M_{tot}}}\log_2\left(1 + \mathrm{SINR}_m\right).
\end{equation}

Next, we derive analytical expressions for the SINR and achievable sum-capacity $R$, given in \eqref{SINR} and \eqref{eq:rate}, respectively, for two different linear precoding techniques, namely, ZF and MRT.

\subsection{Zero-Forcing}
For the ZF precoding scheme, the precoder matrix is constructed using the pseudo-inverse of the estimated effective DL channel matrix as \cite{Marzetta2}
\begin{equation}\label{eq:ZF_precoder}
\mathbf{U}^{\mathrm{ZF}} = \hat{\mathbf{H}}^\mathrm{H}\left(\hat{\mathbf{H}}\hat{\mathbf{H}}^\mathrm{H}\right)^{-1}.
\end{equation}
Next, based on \eqref{eq:beta}, the normalization scalar ${\beta}^{\mathrm{ZF}}$ reads \cite{Marzetta2}
\begin{equation}\label{eq:beta_ZF}
{\beta}^{\mathrm{ZF}} = \left(\sqrt{\mathbb{E}{\left[{\mathrm{Tr}}\left( {{{\left( {{\hat{\mathbf{H}}}{{\hat{\mathbf{H}}}^\mathrm{H}}} \right)}^{-1}}} \right)\right]}}\right)^{-1} = \sqrt{\frac{\left(N-{M_{tot}}\right)\tau_u \rho_u}{{M_{tot}} \left(\tau_u \rho_u + 1\right)}},
\end{equation}
and based on \eqref{eq:received2}, the useful signal term for the detection at the {$m$-th antenna at the UE side} is
\begin{equation} \label{ZF useful}
\sqrt{\rho_d}{\beta}^{\mathrm{ZF}}\mathbb{E}\left[{\mathbf{h}_m^\mathrm{T}{{\mathbf{u}}}_{m}^{\mathrm{ZF}}}\right]s{_m} = \sqrt{\rho_d}{\beta}^{\mathrm{ZF}}s{_m}.
\end{equation}

By substituting \eqref{ZF useful} into \eqref{eq:SI} and \eqref{SINR}, the effective SINR at the {$m$-th antenna in the UE side} for ZF precoding can be written as
\begin{equation} \label{ZF SINR}
\mathrm{SINR}^\mathrm{ZF}_m = \frac{N-{M_{tot}}}{{M_{tot}}} \times \frac{\tau_u \rho_u \rho_d}{{{I}_{\mathrm{RC}}^{\mathrm{ZF}}} + {I}_{\mathrm{NRC},m}^{\mathrm{ZF}}},
\end{equation}
where ${I}_{\mathrm{RC}}^{\mathrm{ZF}} = \rho_d + \tau_u \rho_u + 1$ is the interference plus noise power under ideal reciprocal channel (no NRC), whereas ${I}_{\mathrm{NRC},m}^{\mathrm{ZF}}$ denotes the additional interference power due to NRC, which can be explicitly written as{
\begin{equation} \label{eq:I_ZF_NRC}
\begin{aligned}
{I}_{\mathrm{NRC},m}^{\mathrm{ZF}} &\approx \rho_d \left[\left(1 + \frac{N-M_{tot}}{M_{tot}} \tau_u \rho_u\right) \mathrm{Tr}\left(\mathbf{R}_{\mathbf{a}^\prime_{m}}\right) + \frac{\tau_u \rho_u}{M_{tot}}\left(\mathrm{Tr}\left(\mathbf{R}_{\mathbf{a}^\prime_{m}}\right) - \sigma_{{a}^\prime_{mm}}^2\right) \right.\\
&+\left. \frac{\tau_u \rho_u}{N M_{tot}}\left(1 + \mathrm{Tr}\left(\mathbf{R}_{\mathbf{a}^\prime_{m}}\right)\right) \mathrm{Sum}\left(\mathbf{R}_{\mathrm{c}^\prime_{d}}\right) \right.\\
&+\left. \left[\frac{\tau_u \rho_u + 1}{N}\left(1 + \mathrm{Tr}\left(\mathbf{R}_{\mathbf{a}^\prime_{m}}\right)\right) - \frac{\tau_u \rho_u}{N M_{tot}}\left(\mathrm{Tr}\left(\mathbf{R}_{\mathbf{a}^\prime_{m}}\right) - \sigma_{{a}^\prime_{mm}}^2\right)\right] \left(\mathrm{Tr}\left(\mathbf{R}_{\mathrm{c}^\prime_{d}}\right) + \mathrm{Tr}\left(\mathbf{R}_{\mathrm{c}^\prime_{od}}\right)\right) \right].
\end{aligned}
\end{equation}}
\textit{Proof:} See Appendix \ref{App:ZF}.

\begin{table}[!t]
	\caption{Essential Second-Order Statistics of NRC Variables}
	\label{table:var2}
	\centering
	{
		\begin{tabular}{|c|c|}
			\hline
			Variable & Definition\\
			\hline
			\hline
			$\sigma_{{a}^\prime_{mm}}^2$ & \begin{tabular}{@{}c@{}}Variance of the $m$-th diagonal element\\of UE side NRC matrix $\mathbf{A}^\prime$\end{tabular} \\
			\hline
			$\sigma_{{a}^\prime_{od}}^2$ & \begin{tabular}{@{}c@{}}Average of variances of off-diagonal elements\\of UE side NRC matrix $\mathbf{A}^\prime$\end{tabular} \\
			\hline
			$\sigma_{{c}^\prime_{d}}^2$ & \begin{tabular}{@{}c@{}}Average of variances of diagonal elements\\of BS NRC matrix $\mathbf{C}^\prime$\end{tabular} \\
			\hline
			$\delta_{{c}^\prime_{d}}^2$ & \begin{tabular}{@{}c@{}}Average of cross-correlations of diagonal elements\\of BS NRC matrix $\mathbf{C}^\prime$\end{tabular} \\
			\hline
			$\sigma_{{c}^\prime_{od}}^2$ & \begin{tabular}{@{}c@{}}Average of variances of off-diagonal elements\\of BS NRC matrix $\mathbf{C}^\prime$\end{tabular} \\
			\hline
	\end{tabular}}
\end{table}
{Note that based on \eqref{eq:I_ZF_NRC}, the only NRC characteristics that eventually affect the power of interference are $\sigma_{{a}^\prime_{mm}}^2$ (denoting the variance of the $m$-th diagonal element in $\mathbf{A}^\prime$), $\mathrm{Tr}(\mathbf{R}_{\mathbf{a}^\prime_{m}})$ (denoting the sum of variances of all the elements in the corresponding row of $\mathbf{A}^\prime_k$), $\mathrm{Tr}(\mathbf{R}_{\mathrm{c}^\prime_{od}})$ (denoting the sum of variances of off-diagonal elements in $\mathbf{C}^\prime$), $\mathrm{Tr}(\mathbf{R}_{\mathrm{c}^\prime_{d}})$ (denoting the sum of variances of diagonal elements in $\mathbf{C}^\prime$), and $\mathrm{Sum}(\mathbf{R}_{\mathrm{c}^\prime_{d}})$ (denoting the sum of variances and cross-correlations of diagonal elements in $\mathbf{C}^\prime$). Whereas, other statistical quantities, namely, the cross-correlations of diagonal and off-diagonal elements in ${\mathbf{A}^\prime}$ as well as cross-correlations of off-diagonal elements in $\mathbf{C}'$, do not affect the interference power. In general, different entries of the involved NRC covariance matrices (the $\mathbf{R}$ matrices) can have different values. However, since only the sum of the diagonal values or the sum of all the values in the covariance matrices have impact on the interference power, we parameterize these essential NRC characteristics by their average values for notational simplicity. Thus, the essential NRC characteristics which affect the interference power are listed in \tablename{~\ref{table:var2}}.
Note that when these NRC characteristics} are set to {$0$}, then ${I}_{\mathrm{NRC},m}^{\mathrm{ZF}}=0$, and {if further interpreted in the special case of single-antenna UEs}, \eqref{ZF SINR} reduces to the SINR expression given in \cite{Marzetta2} for the ideal reciprocal case.

\subsection{Maximum Ratio Transmission}
For the MRT case, the precoder matrix is constructed as \cite{Marzetta1}
\begin{equation}\label{eq:MRT_precoder}
{\mathbf{U}}^{\mathrm{MRT}} = \hat{\mathbf{H}}^\mathrm{H}.
\end{equation}
Therefore, based on \eqref{eq:beta}, the normalization scalar ${\beta}^{\mathrm{MRT}}$ reads \cite{Marzetta1}
\begin{equation}\label{eq:beta_MRT}
{\beta}^{\mathrm{MRT}} = \left(\sqrt{\mathbb{E}{\left[{\mathrm{Tr}} {\left( {{\hat{\mathbf{H}}}{{\hat{\mathbf{H}}}^\mathrm{H}}} \right)}\right]}}\right)^{-1} = \sqrt{\frac{\tau_u \rho_u + 1}{N {M_{tot}} \tau_u \rho_u}}.
\end{equation}
Based on \eqref{eq:received2}, the useful signal term for the detection at the {$m$-th antenna in the UE side} is
\begin{equation} \label{MRT useful}
\sqrt{\rho_d}{\beta}^{\mathrm{MRT}}\mathbb{E}\left[{\mathbf{h}_m^\mathrm{T}{{\mathbf{u}}}_{m}^{\mathrm{MRT}}}\right]s_{{m}} = \sqrt{\rho_d}{\beta}^{\mathrm{MRT}}{\frac{N \tau_u \rho_u}{\tau_u \rho_u + 1}}s{_m}.
\end{equation}

Stemming from this, the effective SINR at the {$m$-th antenna in the UE side}, defined in \eqref{SINR} can now be expressed as
\begin{equation} \label{MRT SINR}
\mathrm{SINR}^\mathrm{MRT}_m = \frac{N}{{M_{tot}}} \times \frac{\tau_u \rho_u \rho_d}{{{I}_{\mathrm{RC}}^{\mathrm{MRT}}} + {I}_{\mathrm{NRC},m}^{\mathrm{MRT}}},
\end{equation}
where ${I}_{\mathrm{RC}}^{\mathrm{MRT}} = \left(\rho_d + 1\right) \left(\tau_u \rho_u + 1\right)$ is the interference and noise power under reciprocal channel, whereas ${I}_{\mathrm{NRC},m}^{\mathrm{MRT}}$ denotes the additional interference power due to NRC, and can be explicitly written as
\begin{equation} \label{eq:I_MRT_NRC}
\begin{aligned}
{I}_{\mathrm{NRC},m}^{\mathrm{MRT}} &= \rho_d \left[\left(1 + \frac{N+M_{tot}}{M_{tot}} \tau_u \rho_u\right) \mathrm{Tr}\left(\mathbf{R}_{\mathbf{a}^\prime_{m}}\right) - \frac{\tau_u \rho_u}{M_{tot}}\left(\mathrm{Tr}\left(\mathbf{R}_{\mathbf{a}^\prime_{m}}\right) - \sigma_{{a}^\prime_{mm}}^2\right) \right.\\
&+\left. \frac{\tau_u \rho_u}{N M_{tot}}\left(1 + \mathrm{Tr}\left(\mathbf{R}_{\mathbf{a}^\prime_{m}}\right)\right) \mathrm{Sum}\left(\mathbf{R}_{\mathrm{c}^\prime_{d}}\right) \right.\\
&+\left. \left[\frac{\tau_u \rho_u + 1}{N}\left(1 + \mathrm{Tr}\left(\mathbf{R}_{\mathbf{a}^\prime_{m}}\right)\right) - \frac{\tau_u \rho_u}{N M_{tot}}\left(\mathrm{Tr}\left(\mathbf{R}_{\mathbf{a}^\prime_{m}}\right) - \sigma_{{a}^\prime_{mm}}^2\right)\right]  \left(\mathrm{Tr}\left(\mathbf{R}_{\mathrm{c}^\prime_{d}}\right) + \mathrm{Tr}\left(\mathbf{R}_{\mathrm{c}^\prime_{od}}\right)\right) \right].
\end{aligned}
\end{equation}
\textit{Proof:} See Appendix \ref{App:MRT}.

With the very same reasoning as in the ZF precoding scenario, the only NRC characteristics which affect the power of interference are the ones listed in \tablename{~\ref{table:var2}}. Thus, when these NRC parameters are set to {$0$}, then ${I}_{\mathrm{NRC},m}^{\mathrm{MRT}}=0$ and {in the single-antenna UE scenario,} \eqref{MRT SINR} reduces again to the SINR expression given in \cite{Marzetta2} for the ideal reciprocal case.

\section{Asymptotic and Non-asymptotic Comparisons and Implications} \label{sec:implications}
In this section, we will address several important implications stemming from the derived {closed-form} SINR and achievable rate expressions. To this end, {both the asymptotic and non-asymptotic} performance behavior of ZF and MRT precoding based systems are first derived and compared. Then, the SINR degradation due to NRC is quantified and analyzed for both precoding techniques.

\subsection{Asymptotic Performance for Large $N$}\label{sec:large_N}
For growing $N$, the previously-derived SINR expressions for ZF and MRT based systems, under NRC, can be shown to be asymptotically identical and have the saturation value
\begin{equation}\label{eq:SNR_N_inf}
\lim\limits_{N\rightarrow\infty} \mathrm{SINR}^\mathrm{ZF}_m = \lim\limits_{N\rightarrow\infty} \mathrm{SINR}^\mathrm{MRT}_m = \frac{1}{\mathrm{Tr}\left(\mathbf{R}_{\mathbf{a}^\prime_{m}}\right) + t_{\mathrm{c}^\prime_{d}}^m \delta_{{c}^\prime_{d}}^2 + t_{\mathrm{c}^\prime_{od}}^m \sigma_{{c}^\prime_{od}}^2},
\end{equation}
where
\begin{equation}
\begin{gathered}
t_{\mathrm{c}^\prime_{d}}^m = 1 + \mathrm{Tr}\left(\mathbf{R}_{\mathbf{a}^\prime_{m}}\right), \\
t_{\mathrm{c}^\prime_{od}}^m = M_{tot}\frac{\tau_u \rho_u + 1}{\tau_u \rho_u} \left(1 + \mathrm{Tr}\left(\mathbf{R}_{\mathbf{a}^\prime_{m}}\right)\right) - \mathrm{Tr}\left(\mathbf{R}_{\mathbf{a}^\prime_{m}}\right) + \sigma_{{a}^\prime_{mm}}^2.
\end{gathered}
\end{equation}
Note that the number of mismatched transceiver chains {and antenna units} increases with the number of antennas which in turn increases the level of interference power due to NRC. Thus, the system is subject to additional interference which cannot be suppressed by NRC-unaware spatial precoders{, even if the number of antennas tends towards infinity}. Therefore, for massive MIMO systems with practical non-reciprocal transceivers {and antenna systems}, the advantage of ZF over MRT in terms of IUI suppression, and hence in SINR performance, reduces and eventually vanishes with increasing number of antennas and transceiver chains. {This is one important finding and will be illustrated also through numerical examples in Section \ref{sec:results}.}

We next quantify the relative achievable rate performance under ZF and MRT precoding schemes with the ratio ${R^\mathrm{ZF}}/{R^\mathrm{MRT}}$, where $R^\mathrm{ZF}$ and $R^\mathrm{MRT}$ are obtained by substituting \eqref{ZF SINR} and \eqref{MRT SINR} into \eqref{eq:rate}, respectively. The asymptotic behavior of this relative achievable rate performance for large number of antennas can be shown to read
\begin{equation}\label{eq:asymp_R}
\lim\limits_{N\rightarrow\infty} \frac{R^\mathrm{ZF}}{R^\mathrm{MRT}} = \lim\limits_{N\rightarrow\infty} \frac{\sum\limits_{m = 1}^{M_{tot}} \log_2 \left(1 + \mathrm{SINR}^\mathrm{ZF}_m\right)}{\sum\limits_{m = 1}^{M_{tot}} \log_2 \left(1 + \mathrm{SINR}^\mathrm{MRT}_m\right)} = 1.
\end{equation}
Based on above, the asymptotic behavior of relative achievable rate under NRC is similar to the reciprocal case presented in \cite{Marzetta2}. However, the implications of these two results are largely different. More specifically, {the combination of \eqref{eq:SNR_N_inf} and \eqref{eq:asymp_R}} establishes that the achievable rates for both precoders have an identical and finite saturation level in the presence of NRC. This saturation level can be expressed in closed-form by substituting the expression in \eqref{eq:SNR_N_inf} to \eqref{eq:rate}. Importantly, even if the UL pilot SNR ($\rho_u$) tends towards infinity, reflecting perfect uplink CSI, the rates saturate to an identical finite level. On the other hand, for an ideal reciprocal channel, {by substituting zeros for all the NRC parameters in the denominator of \eqref{eq:SNR_N_inf}}, the asymptotic result implies that the SINRs, and therefore the rates, grow without bound for both precoding schemes {even under finite UL pilot SNR \cite{massMIMOtut,Marzetta2}. Hence, there is a fundamental difference in the impacts of NRC and UL channel estimation errors}. {These differences will be illustrated through numerical examples in Section \ref{sec:results} and are other important findings of this article.}

\subsection{Non-Asymptotic Comparison of SINR Performance} \label{sec:gen perf}
We next pursue a non-asymptotic comparison of the achievable SINRs {at the $m$-th antenna in the UE side} between ZF and MRT precoding schemes under NRC. Building on the SINR expressions in \eqref{ZF SINR} and \eqref{MRT SINR}, the following relation can be deduced
\begin{equation} \label{eq:gen performance}
\frac{\mathrm{SINR}^\mathrm{ZF}_m}{\mathrm{SINR}^\mathrm{MRT}_m}= 1 + \frac{{M_{tot}}}{N}\left(\mathrm{SINR}^\mathrm{ZF}_m - 1\right) + \left(1 - \frac{{M_{tot}}}{N}\right) \frac{2\rho_d\tau_u\rho_u {\left(\mathrm{Tr}\left(\mathbf{R}_{\mathbf{a}^\prime_{m}}\right) - \left(\mathrm{Tr}\left(\mathbf{R}_{\mathbf{a}^\prime_{m}}\right) - \sigma_{{a}^\prime_{mm}}^2\right)/M_{tot}\right)}}{\rho_d + \tau_u \rho_u + 1 + {I}_{\mathrm{NRC},m}^{\mathrm{ZF}}}.
\end{equation}
Based on above, since $N > {M_{tot}}$, ZF outperforms MRT in the achievable SINR, and consequently in the capacity lower bound, if  $\mathrm{SINR}^\mathrm{ZF}_m \geq 1$. In the special case of $N\rightarrow\infty$, the ratio in \eqref{eq:gen performance} tends towards one, conforming with the previous asymptotic results.

In practical scenarios where the channel non-reciprocity level is not overly high, and considering the high SNR region with reasonably good UL channel estimation accuracy, the SINR is always greater than one for ZF precoding scheme. Therefore, in the high SNR region, \eqref{eq:gen performance} shows that ZF has better non-asymptotic performance compared to MRT. On the other hand, in the low SNR region, the performance of both systems are limited by noise and the difference becomes negligible.

\subsection{SINR Degradation at High SNR}
In order to quantify the SINR degradation under non-reciprocal channels with respect to ideal reciprocal channel reference case, we define the metric
\begin{equation} \label{eq:SINR deg}
\alpha = \frac{\mathrm{SINR}_{\mathrm{RC}} - \mathrm{SINR}_{\mathrm{NRC}}}{\mathrm{SINR}_{\mathrm{RC}}}.
\end{equation}
In \eqref{eq:SINR deg}, $\mathrm{SINR}_{\mathrm{NRC}}$ stands for the SINR with non-reciprocal channels calculated based on \eqref{SINR} and for which closed-form analytic expressions are given in \eqref{ZF SINR} and \eqref{MRT SINR} under ZF and MRT precoding schemes, respectively. Furthermore, $\mathrm{SINR}_{\mathrm{RC}}$ denotes the SINR with reciprocal channels for which closed-form expressions can be obtained under ZF and MRT precoding schemes from \cite{Marzetta2} {for the single-antenna UE scenario}, or by setting the NRC parameters to $0$ in \eqref{ZF SINR} and \eqref{MRT SINR}, respectively{, in a more general case}. To compare the relative SINR degradation of ZF and MRT precoding schemes, we also define the ratio $\alpha^{\mathrm{ZF}/\mathrm{MRT}} = \frac{\alpha^\mathrm{ZF}}{\alpha^\mathrm{MRT}}$, where ${\alpha^\mathrm{ZF}}$ and ${\alpha^\mathrm{MRT}}$ are calculated using \eqref{eq:SINR deg} with their corresponding $\mathrm{SINR}_{\mathrm{NRC}}$ and $\mathrm{SINR}_{\mathrm{RC}}$ expressions for ZF and MRT precoding schemes, respectively.

In the high SNR region, when $\rho_d \gg 1$, this ratio {for the $m$-th antenna in the UE side} can be shown to read
\begin{equation} \label{eq:deg}
\lim\limits_{\rho_d\rightarrow\infty} \frac{\alpha^\mathrm{ZF}_m}{\alpha^\mathrm{MRT}_m} \overset{\Delta}{=} \alpha^{\mathrm{ZF}/\mathrm{MRT}}_{\infty,m} = \frac{I_0 + \tau_u\rho_u{I}_{\mathrm{NRC},m}^{\mathrm{ZF}}/\rho_d}{I_0 + 2\tau_u\rho_u{\left(\mathrm{Tr}\left(\mathbf{R}_{\mathbf{a}^\prime_{m}}\right) - \left(\mathrm{Tr}\left(\mathbf{R}_{\mathbf{a}^\prime_{m}}\right) - \sigma_{{a}^\prime_{mm}}^2\right)/M_{tot}\right)}},
\end{equation}
where
\begin{equation}
I_0 = \left(2\tau_u\rho_u{\left(\mathrm{Tr}\left(\mathbf{R}_{\mathbf{a}^\prime_{m}}\right) - \left(\mathrm{Tr}\left(\mathbf{R}_{\mathbf{a}^\prime_{m}}\right) - \sigma_{{a}^\prime_{mm}}^2\right)/M_{tot}\right)} + {I}_{\mathrm{NRC},m}^{\mathrm{ZF}}/\rho_d + 1\right) {I}_{\mathrm{NRC},m}^{\mathrm{ZF}}/\rho_d.
\end{equation}
From \eqref{eq:deg}, it can be seen that $\alpha^{\mathrm{ZF}/\mathrm{MRT}}_{\infty,m} > 1$ when ${I}_{\mathrm{NRC},m}^{\mathrm{ZF}}/\rho_d > 2{\left(\mathrm{Tr}\left(\mathbf{R}_{\mathbf{a}^\prime_{m}}\right) - \left(\mathrm{Tr}\left(\mathbf{R}_{\mathbf{a}^\prime_{m}}\right) - \sigma_{{a}^\prime_{mm}}^2\right)/M_{tot}\right)}$, implying that ZF precoding is more sensitive to channel non-reciprocity than MRT, that is, the SINR degradation due to NRC is higher for ZF than for MRT, at large SNR. This is intuitive as the ZF based interference suppression requires accurate channel knowledge. Note that based on \eqref{eq:I_ZF_NRC}, for practical setting of $\tau_u \geq {M_{tot}}$, this holds when 
\begin{equation}\label{eq:extra}
\rho_u > \frac{1}{N-{M_{tot}}}.
\end{equation}
This is because when $N \gg M_{tot}$, the inequality given in \eqref{eq:extra} boils down to $\rho_u >1/N$, which will be satisfied, in general, for all practical values of $\rho_u$.

\section{Numerical Results, Implications and Discussion} \label{sec:results}
In this section, we provide extensive numerical evaluations of the derived analytical SINR and achievable rate expressions for precoded multi-user massive MIMO system under NRC and imperfect CSI. We also study the behavior of the {DL} system spectral efficiency, defined as \cite{Marzetta2}
\begin{equation}\label{eq:spec_eff}
\eta_s = \left(1 - \frac{\tau_u}{T}\right) R = \left(1 - \frac{\tau_u}{T}\right) {\sum\limits_{m = 1}^{M_{tot}}}\log_2\left(1 + \mathrm{SINR}_m\right),
\end{equation}
where $T$ {refers to the channel coherence interval measured in number of symbols}. Finally, we will discuss and summarize the novel findings of this work based on the derived analytical expressions and obtained numerical results.

\subsection{Obtained Numerical Results}
The baseline evaluation scenario consists of a BS which is equipped with $N = 100$ antenna elements and {either single-antenna, dual-antenna or 4-antenna UEs, with a total of $M_{tot} = 20$ antennas,} that are served simultaneously through either ZF or MRT precoding. We assume that the channel coherence time is $1$ms, which corresponds to one radio sub-frame in $3$GPP LTE/LTE-Advanced radio network \cite{T196} and specifically each coherence interval contains $T = 196$ symbols, while the number of {UE antenna-specific} UL pilots is always equal to the {total} number of {the UE side antennas}, i.e., $\tau_u = {M_{tot}}$. The UL SNR is set to $\rho_u = 0$ dB, while DL SNR is chosen to be $\rho_d = 20$ dB. These are the baseline simulation settings, while some of the parameter values are also varied in the evaluations.

In the simulations, NRC matrices $\mathbf{A}$ and $\mathbf{C}$ are generated based on $\mathbf{A}^\prime$ and $\mathbf{C}^\prime$ since $\mathbf{A} = \mathbf{I}_{M_{tot}} + \mathbf{A}^\prime$ and $\mathbf{C} = \mathbf{I}_{N} + \mathbf{C}^\prime$. As shown in Section \ref{sec:analysis} and \tablename{~\ref{table:var2}}, only the averages of certain variance and cross-correlation values affect the performance, while in principle the individual values could all be different. However, for simulation simplicity, we assume that all the individual entries of the involved second-order statistics, i.e. those listed in \tablename{~\ref{table:var2}}, are the same as their average values. We also assume that $\sigma_{{a}^\prime_{mm}}^2$ is the same for all the values of $m$ and is equal to $\sigma_{{a}^\prime_{d}}^2$. Thus, for each realization, the block-diagonal matrix $\mathbf{A}^\prime$ is generated based on $\mathbf{A}^\prime_k$ in which the diagonal entries are generated as i.i.d. $\mathcal{CN}\left(0,\sigma_{{a}^\prime_{d}}^2\right)$ whereas off-diagonal entries are i.i.d. $\mathcal{CN}\left(0,\sigma_{{a}^\prime_{od}}^2\right)$. Similarly, the off-diagonal entries of $\mathbf{C}^\prime$ are generated as i.i.d. $\mathcal{CN}\left(0,\sigma_{{c}^\prime_{od}}^2\right)$, while the diagonal entries have Gaussian distribution with zero mean and variance $\sigma_{{c}^\prime_{d}}^2$ and cross-correlation $\delta_{{c}^\prime_{d}}^2$. Kindly note that Gaussian distribution is chosen only as an example for the simulation and evaluation simplicity while the provided results apply to any distribution with the same variance and cross-correlation values. Also note that the independence assumption applies only to the entries whose cross-correlations do not have any impact on the system performance.

\begin{figure}[!t]
	\centering
	\subfloat[]{\includegraphics[height=0.26\textheight]{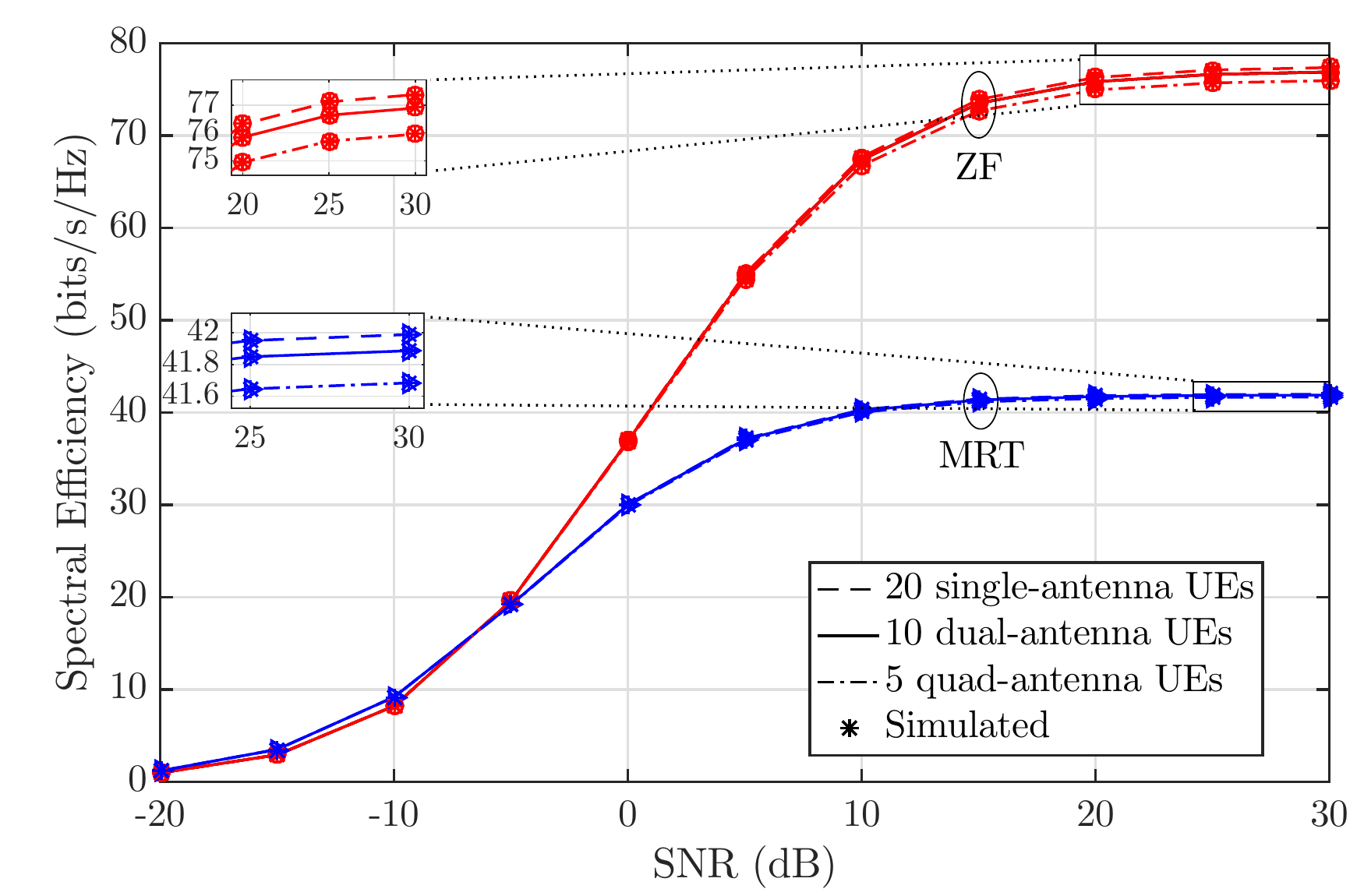}
		\label{subfig:SE_SNR}}\quad
	\subfloat[]{\includegraphics[height=0.26\textheight]{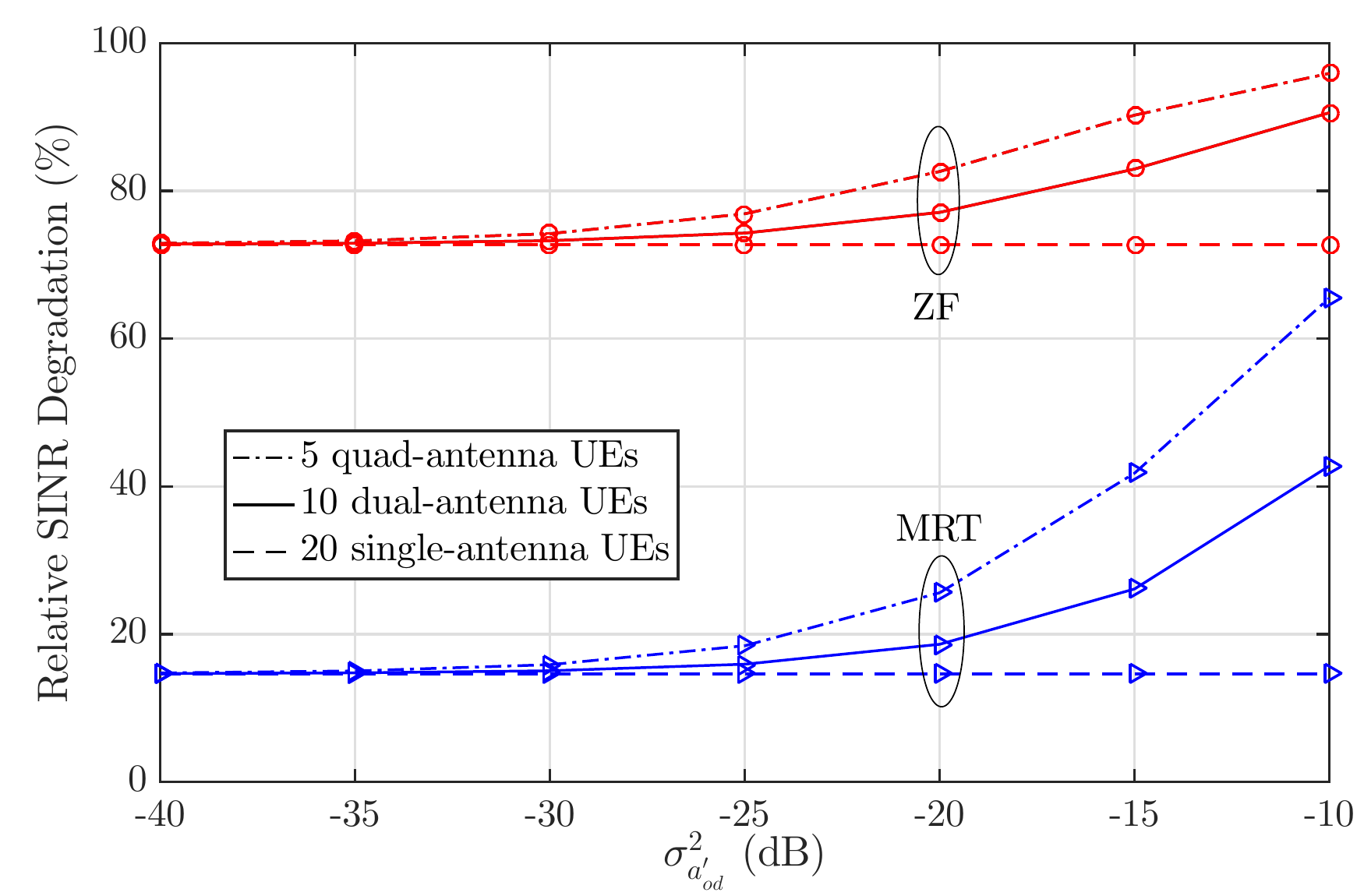}
		\label{subfig:SINRdeg_MCue}}
	\caption{{\protect\subref{subfig:SE_SNR} System spectral efficiency vs. DL SNR ($\rho_d$) with $\sigma_{{a}^\prime_{od}}^2 = -30$ dB, and \protect\subref{subfig:SINRdeg_MCue} relative SINR degradation ($\alpha$) vs. $\sigma_{{a}^\prime_{od}}^2$ with $\rho_d = 20$ dB, for $N = 100$, $M_{tot} = 20$, $\tau_u = M_{tot}$, $\rho_u = 0$ dB, $T = 196$, $\sigma_{{a}^\prime_{d}}^2 = -20$ dB, $\sigma_{{c}^\prime_{d}}^2 = -20$ dB, $\sigma_{{c}^\prime_{od}}^2 = -30$ dB, and $\delta_{{c}^\prime_{d}}^2 = -30$ dB.}}
	\label{fig:UEantennas}
\end{figure}
In \figurename{~\ref{fig:UEantennas}}{(a)}, the system spectral efficiency is plotted against DL SNR for different number of antennas in each UE, while the total number of antennas in the UE side is fixed at $M_{tot} = 20$. In obtaining the curves, the derived analytical expressions in \eqref{ZF SINR} and \eqref{MRT SINR} are plugged into \eqref{eq:spec_eff} for ZF and MRT precoding schemes, respectively. In addition to that, simulated points are obtained via extensive empirical SINR and corresponding spectral efficiency evaluations, without any approximations, which are averaged over $1000$ independent channel and NRC variable realizations. As can be seen, when the total number of antennas in the UE side is fixed, the spectral efficiency of the system is slightly higher in networks with lower number of antennas in each UE. {This can be understood based on \figurename{~\ref{fig:UEantennas}}{(b)} where the relative SINR degradation is plotted against $\sigma_{{a}^\prime_{od}}^2$, which is an indicator of mutual coupling mismatch variance between the antenna elements of each individual UE (in dual- and quad-antenna UE cases). As illustrated, the number of antennas in each UE has essentially no impact on the performance when UE side mutual coupling mismatch level is small. Whereas, for higher values of $\sigma_{{a}^\prime_{od}}^2$, the relative SINR degradation is already clearly higher in the scenarios with higher number of antennas per UE and the difference gets larger as $\sigma_{{a}^\prime_{od}}^2$ increases}. However, {as shown in \figurename{~\ref{fig:UEantennas}}{(a)},} even for relatively poorly NRC calibrated scenario (high {practical} NRC parameter values, e.g., $\sigma_{{a}^\prime_{d}}^2 = \sigma_{{c}^\prime_{d}}^2 = -20$ dB and $\sigma_{{a}^\prime_{od}}^2 = \sigma_{{c}^\prime_{od}}^2 = \delta_{{c}^\prime_{d}}^2 = -30$ dB), the difference between single-antenna and multi-antenna UE scenarios is very small. Therefore, in the continuation, we focus on single-antenna UE scenario which is commonly of highest interest in massive MIMO literature \cite{massMimo0,massMIMO01,Marzetta0,massMIMOeffic,Marzetta2,massMIMOchannel,TDDprefer,ourGlobecom14,idealnonRecip1,idealnonRecip2,idealnonRecip3,nonRecipinMassMIMO,Marzetta1,Marzetta4}.

\begin{figure}[!t]
	\centering
		\includegraphics[height=0.26\textheight]{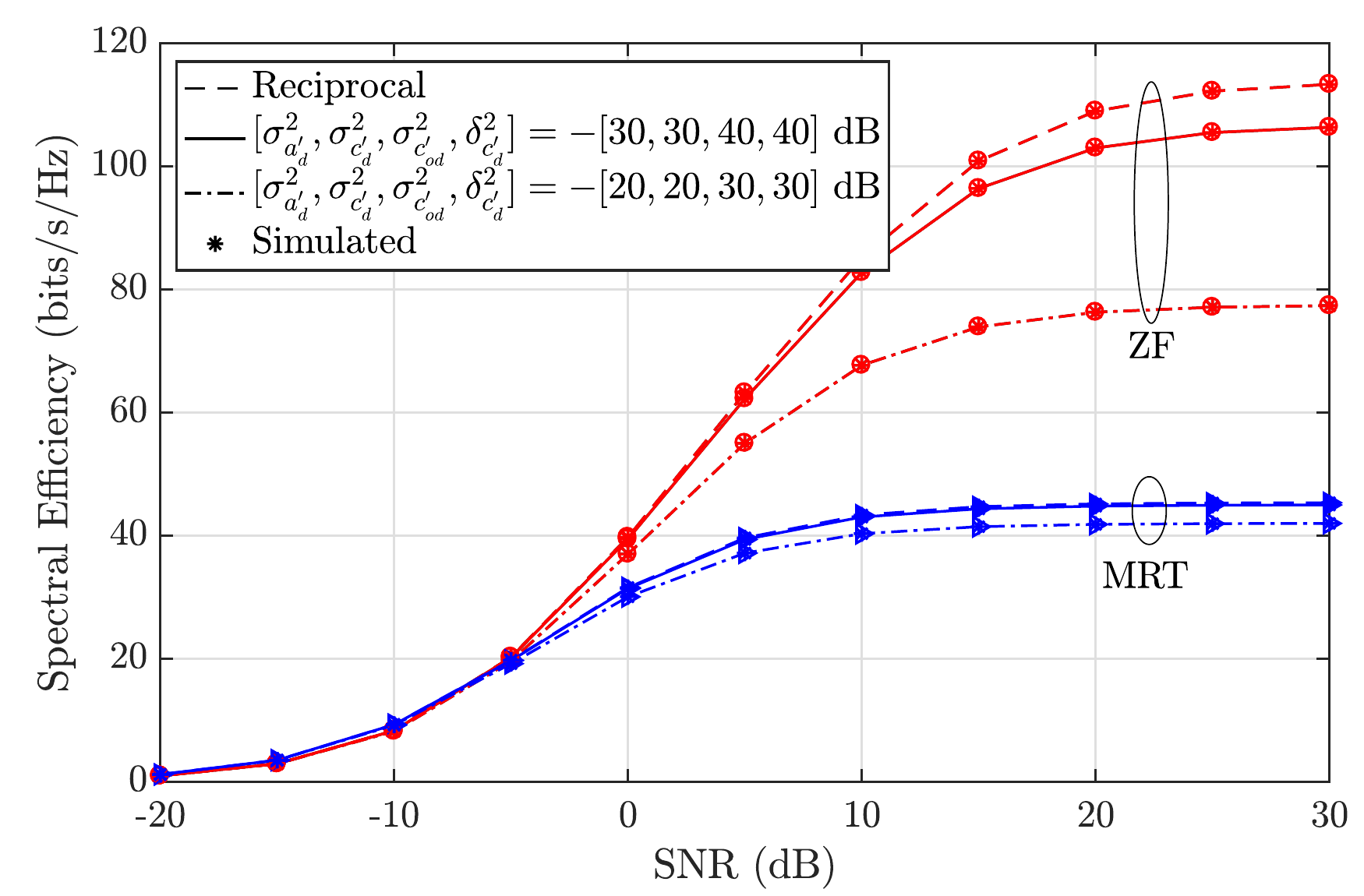}
		\caption{System spectral efficiency vs. DL SNR ($\rho_d$) for $N = 100$, {$M_{tot} = 20$, $K = 20$}, $\tau_u = {M_{tot}}$, $\rho_u = 0$ dB, $T = 196$.}
		\label{fig:SNR1}
\end{figure}
\begin{figure}[!t]
		\centering
		\includegraphics[height=0.26\textheight]{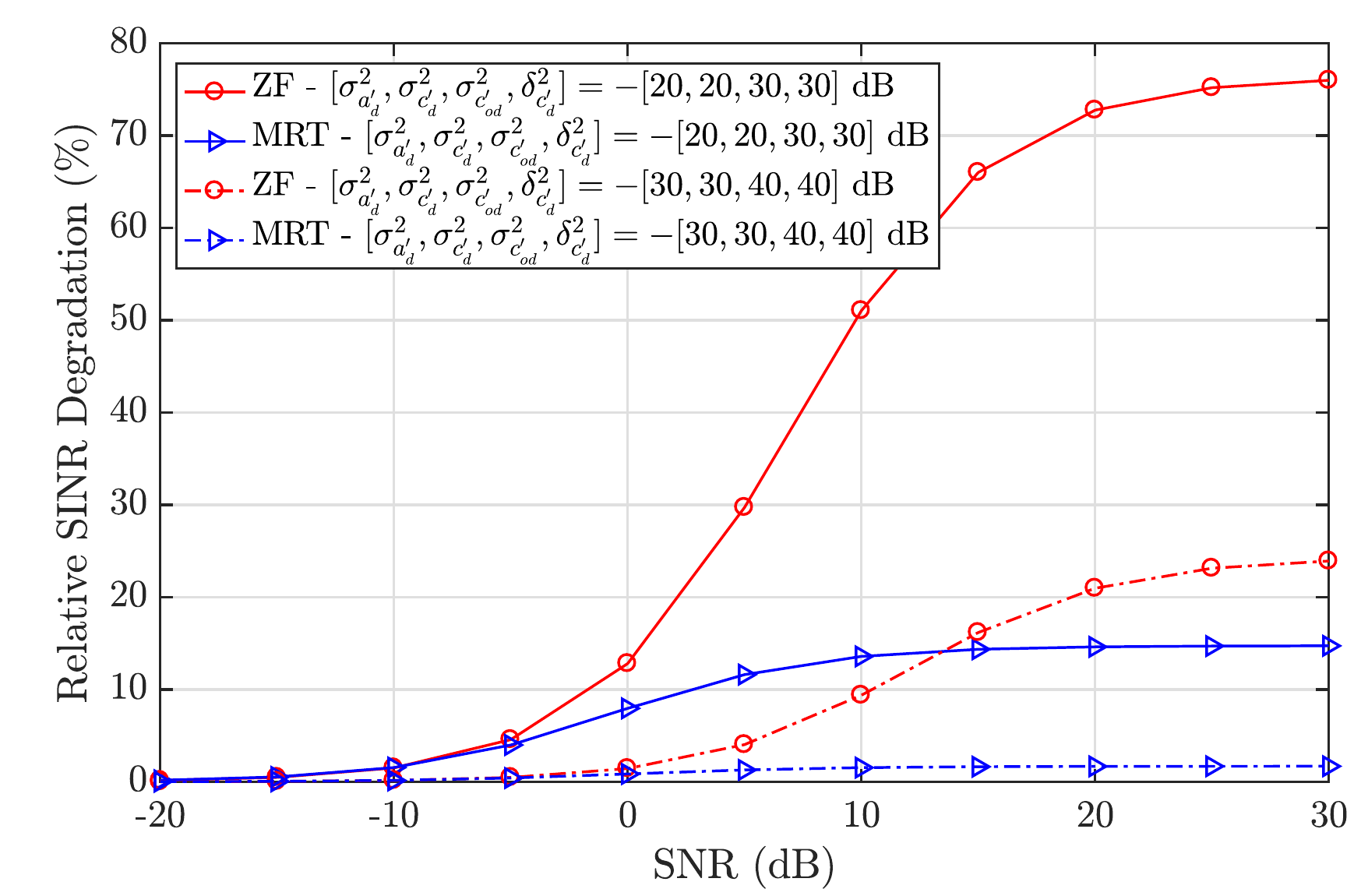}
		\caption{Relative SINR degradation ($\alpha$) vs. DL SNR ($\rho_d$) for $N = 100$, {$M_{tot} = 20$, $K = 20$}, $\tau_u = {M_{tot}}$, $\rho_u = 0$ dB, $T = 196$.}
		\label{fig:SNR2}
\end{figure}
In \figurename{~\ref{fig:SNR1}} and \figurename{~\ref{fig:SNR2}}, the spectral efficiency and relative SINR degradation curves are plotted against DL SNR for indicated NRC parameter settings. {Simulated curves in \figurename{~\ref{fig:SNR1}} are similarly obtained via extensive empirical evaluations by averaging $1000$ independent channel and NRC realizations. In general, as can be seen in \figurename{~\ref{fig:UEantennas}} and \figurename{~\ref{fig:SNR1}},} the analytical and simulated curves for both ZF and MRT have a perfect match evidencing the excellent accuracy of derived expressions despite the involved approximations. Thus, in the continuation we will use only the derived analytical expressions. As illustrated in \figurename{~\ref{fig:SNR1}} and \figurename{~\ref{fig:SNR2}}, in low SNR region, the effect of channel non-reciprocity on both precoding schemes is negligible as the performance is limited by noise. On the other hand, in high SNR region, there is a substantial performance loss, especially for ZF precoding scheme. For instance, from \figurename{~\ref{fig:SNR1}} we can observe that for ZF at $\rho_d = 15$ dB, when the system is subject to {relatively low-quality NRC calibration ($\sigma_{{a}^\prime_{d}}^2 = \sigma_{{c}^\prime_{d}}^2 = -20$ dB and $\sigma_{{c}^\prime_{od}}^2 = \delta_{{c}^\prime_{d}}^2 = -30$ dB)}, the system spectral efficiency has decreased by {$27$} bits/s/Hz compared to the fully reciprocal channel case. For the same settings, the degradation for MRT precoding scheme is only {$3$} bits/s/Hz showing that MRT is {substantially} less sensitive to channel non-reciprocity compared to ZF. 

\begin{figure}[!t]
	\centering
	\subfloat[]{\includegraphics[height=0.26\textheight]{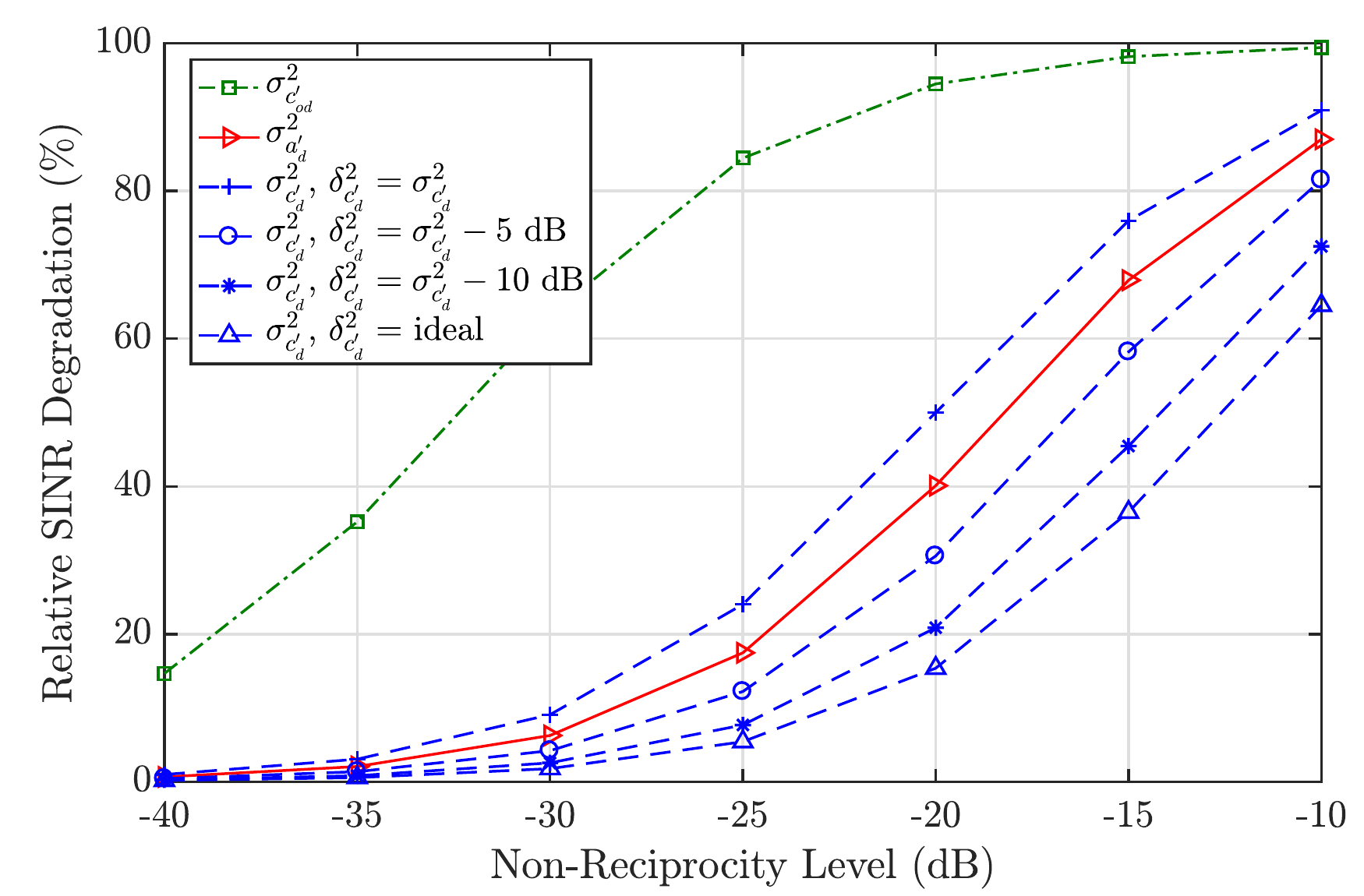}
		\label{subfig:ZF}}\quad
	\subfloat[]{\includegraphics[height=0.26\textheight]{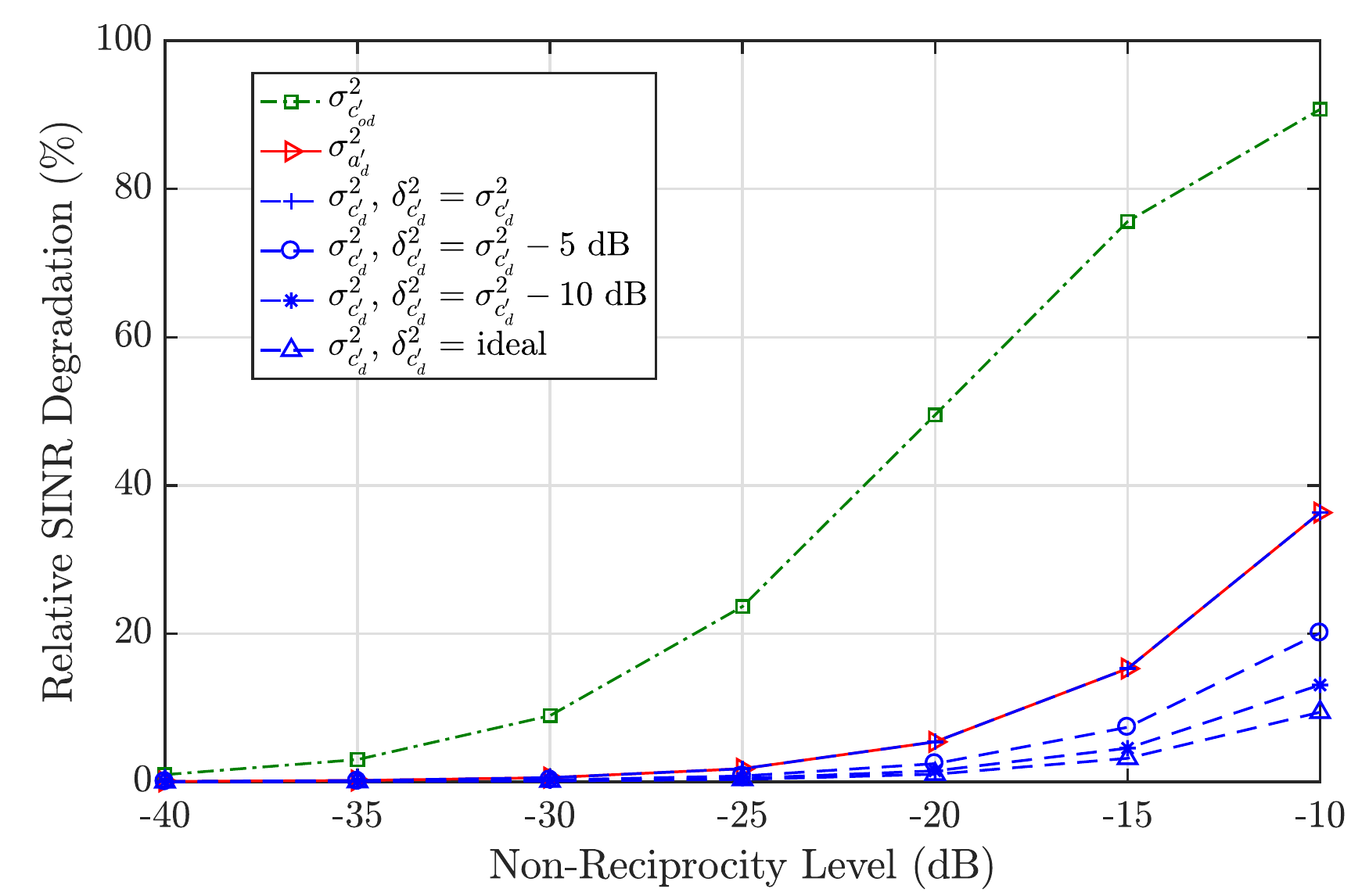}
		\label{subfig:MRT}}
	\caption{Relative SINR degradation ($\alpha$) {in \protect\subref{subfig:ZF} ZF and \protect\subref{subfig:MRT} MRT precoded systems} vs. the level of an individual non-reciprocity source (with others being zero, i.e., ideal) for $N = 100$, {$M_{tot} = 20$, $K = 20$}, $\rho_d = 20$, $\tau_u = {M_{tot}}$, $\rho_u = 0$ dB, $T = 196$.}
	\label{fig:NRC}
\end{figure}
Based on the derived expressions in \eqref{eq:I_ZF_NRC} and \eqref{eq:I_MRT_NRC}, the contributions of {$\sigma_{{c}^\prime_{od}}^2$, $\sigma_{{a}^\prime_{d}}^2$, $\delta_{{c}^\prime_{d}}^2$, and $\sigma_{{c}^\prime_{d}}^2$} to the total received interference are proportional to {$N \times M_{tot}$, $N$, $N$, and $M_{tot}$}, respectively. In typical settings with $N \gg {M_{tot}}$ (which is also the case with $N = 100$ and {$M_{tot} = 20$}), these cofactors satisfy the relation {$N \times M_{tot} > {N} > M_{tot}$}, and hence the system has the highest sensitivity with respect to {$\sigma_{{c}^\prime_{od}}^2$} and the lowest sensitivity with respect to {$\sigma_{{c}^\prime_{d}}^2$}. {This is one of the important technical implications, relevant in practical large-array system deployments and NRC calibration algorithm development, that are stemming from this work.} In order to demonstrate this effect, in \figurename{~\ref{fig:NRC}}, the relative SINR degradation is plotted against different levels of each channel non-reciprocity parameter individually, i.e., when the level of one channel non-reciprocity parameter is varied, all other channel non-reciprocity parameter values are deliberately set to $0$. {Note that, in order to better demonstrate the impacts of $\delta_{{c}^\prime_{d}}^2$ on the SINR degradation, the effects of $\sigma_{{c}^\prime_{d}}^2$ and $\delta_{{c}^\prime_{d}}^2$ are grouped together, since the level of cross-correlation between elements in $\mathbf{c^\prime_{d}}$, $\delta_{{c}^\prime_{d}}^2$, is always upper-bounded by the corresponding variances of those elements, $\sigma_{{c}^\prime_{d}}^2$. The effects of both $\sigma_{{c}^\prime_{d}}^2$ and $\delta_{{c}^\prime_{d}}^2$ can be distinguished by the offset chosen between these two variables which ranges from $\delta_{{c}^\prime_{d}}^2 = \sigma_{{c}^\prime_{d}}^2$ to $\delta_{{c}^\prime_{d}}^2 = 0$.} As expected, the obtained results show that both ZF and MRT precoding schemes are most sensitive to the {variance} of the off-diagonal elements of the BS non-reciprocity matrix. For instance, for the case with ${\sigma_{{c}^\prime_{od}}^2} = -25$ dB, the SINR degradation is approximately {$85$}\% for ZF and {$25$}\% for MRT{, which will be mapped to $42$\% and $13$\% of spectral efficiency degradation, respectively}. The SINR degradation is, in turn, the least sensitive against the {variance} of the diagonal entries of BS side non-reciprocity matrix. It is seen that {when $\delta_{{c}^\prime_{d}}^2 = 0$,} ZF precoded system starts to have observable performance loss{, i.e., the SINR degradation is more than $10$\%,} for values of {$\sigma_{{c}^\prime_{d}}^2 > -23$} dB, whereas for MRT precoded system this threshold value is as high as {$\sigma_{{c}^\prime_{d}}^2 > -10$} dB. The sensitivity with respect to {the variance of diagonal elements in the} UE side {NRC matrix} {and the cross-correlations between diagonal elements in the BS side NRC matrix are} also considerably high especially for ZF precoding. For instance, the SINR degradation increases from {$17$}\% to {$40$}\%, when {$\sigma_{{a}^\prime_{d}}^2$} is increased from $-25$ dB to $-20$ dB, and from $24$\% to $50$\%, when $\delta_{{c}^\prime_{d}}^2$ and $\sigma_{{c}^\prime_{d}}^2$ are jointly increased from $-25$ dB to $-20$ dB.

\begin{figure}[!t]
		\centering
		\includegraphics[height=0.26\textheight]{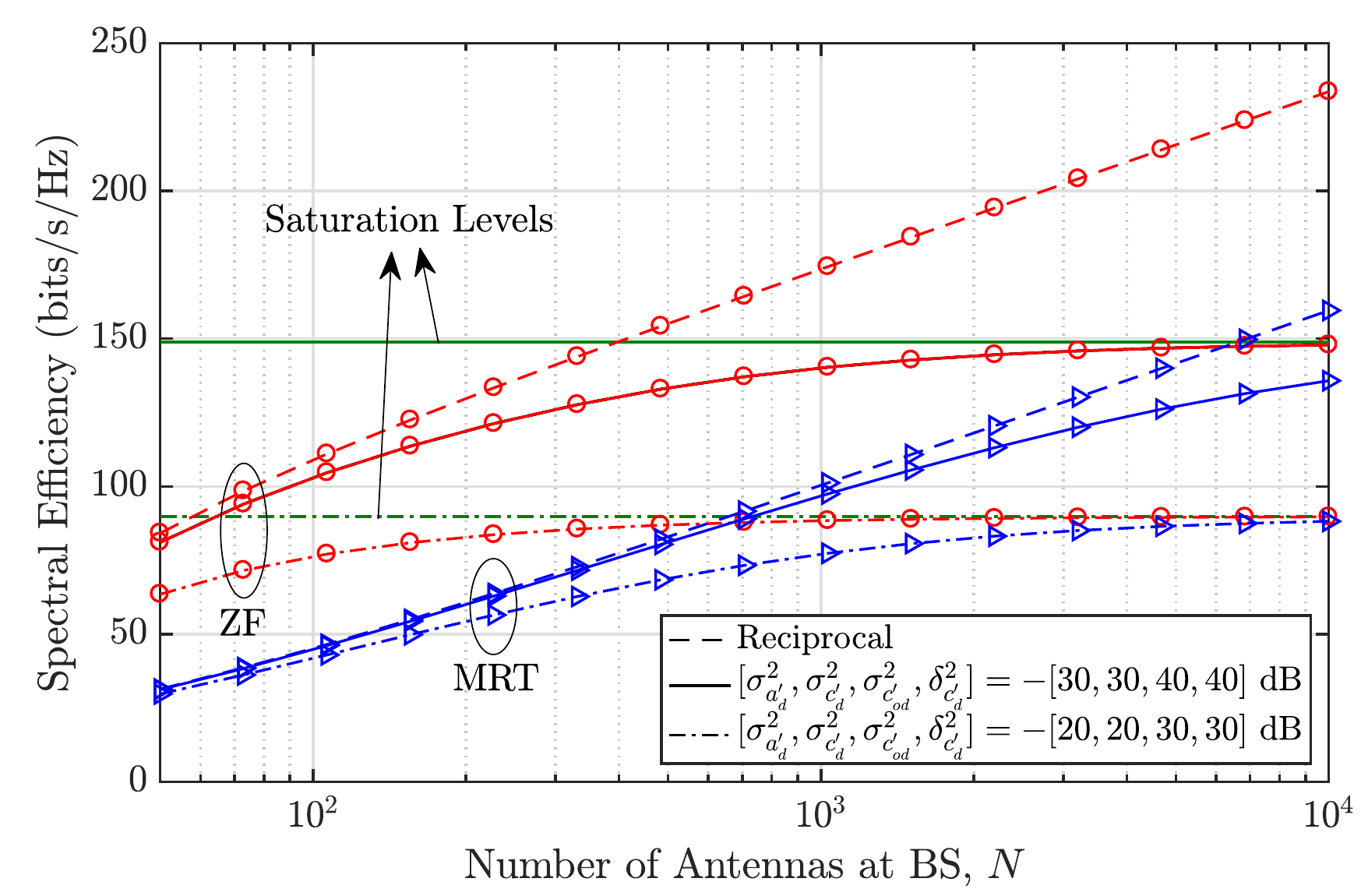}
		\caption{System spectral efficiency vs. the number of antennas at BS ($N$) for {$M_{tot} = 20$, $K = 20$}, $\rho_d = 20$, $\tau_u = {M_{tot}}$, $\rho_u = 0$ dB, $T = 196$. Saturation levels based on \eqref{eq:SNR_N_inf} are plotted in green horizontal lines for the two indicated NRC parameter settings.}
		\label{fig:N1}
\end{figure}
\begin{figure}[!t]
		\centering
		\includegraphics[height=0.26\textheight]{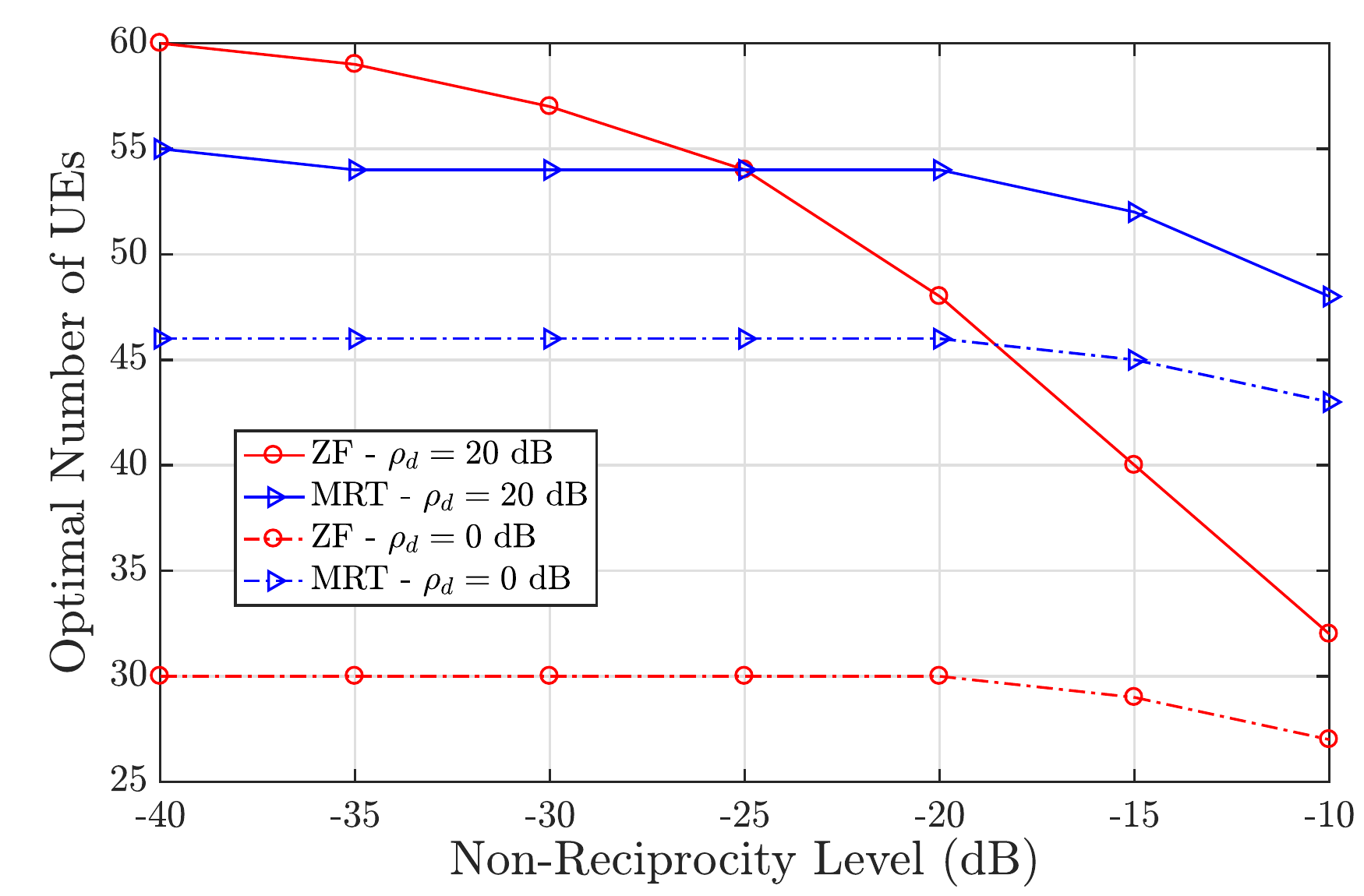}
		\caption{Optimal number of {single-antenna} UEs to maximize system spectral efficiency vs. non-reciprocity level ({$\sigma_{{a}^\prime_{d}}^2 = \sigma_{{c}^\prime_{d}}^2 = $} NRC level, while {$\sigma_{{c}^\prime_{od}}^2 = \delta_{{c}^\prime_{d}}^2 = $} NRC level $ - 10$ dB) for $N = 100$, {$M_{tot} = 20$, $K = 20$}, $\tau_u = {M_{tot}}$, $\rho_u = 0$ dB, $T = 196$.}
		\label{fig:Kopt}
\end{figure}
The analytical expressions for the asymptotic achievable performance, derived in Section \ref{sec:implications}, indicated two new results and findings which differ from the ordinary reciprocal channel case; 1) there is a finite saturation level for both MRT and ZF precoding schemes, and 2) this saturation level is identical for both precoding techniques. In order to verify and demonstrate this behavior, the spectral efficiency is plotted against the number of BS antennas in \figurename{~\ref{fig:N1}}. It can be clearly seen that both MRT and ZF spectral efficiency curves indeed saturate towards the levels predicted by the derived analytical expression in \eqref{eq:SNR_N_inf}. As discussed earlier in Section \ref{sec:large_N}, the system is subject to increasing levels of SI and ISI with increasing number of antennas and corresponding mismatched transceiver chains. Since this interference cannot be suppressed by NRC-unaware spatial precoders, in contrast to the reciprocal case, the advantage of ZF over MRT in terms of inter-user interference suppression and higher achievable rates gradually vanishes. It is also important to note that these saturation levels are of large practical relevance since the NRC-induced saturation occurs already with antenna numbers in the order of $10^3$ or even below, while the saturation levels caused, e.g., by pilot contamination often requires $10^5$ antennas to be approached \cite{Emil_recip_rate_limit}.

\figurename{~\ref{fig:Kopt}} shows the impact of channel non-reciprocity on the optimal number of simultaneously scheduled {single-antenna} UEs, $K_{opt}$, to achieve maximal spectral efficiency for two different values of DL SNR, namely, $\rho_d = 20$ dB, $0$ dB. This optimum number is achieved by evaluating \eqref{eq:rate} for all the values of $K$ in the range $N \geq K \geq 1$, and choosing the one which maximizes the spectral efficiency {while the number of antennas in each UE is assumed to be one}. The optimal number of {single-antenna} UEs drops for both precoding techniques as the system is subject to increasing interference power with increasing non-reciprocity levels. In the low SNR regime ($0$ dB), this drop is not severe as the thermal noise has dominating impact on system performance. However, in the high SNR regime ($20$ dB), there is a significant drop in the optimal number of {single-antenna} UEs for ZF, even for moderate channel non-reciprocity levels, say $-30$ dB $< {\sigma_{{a}^\prime_{d}}^2} < -20$ dB, whereas for MRT there is a drop only at fairly severe non-reciprocity levels, e.g., ${\sigma_{{a}^\prime_{d}}^2} > -15$ dB. An interesting and new observation is that, in contrast to high SNR regime behavior in the ordinary reciprocal case, the optimal number of UEs for MRT is higher than that of ZF under moderate channel non-reciprocity levels.

\begin{figure}[t]
	\centering
	\includegraphics[height=0.26\textheight]{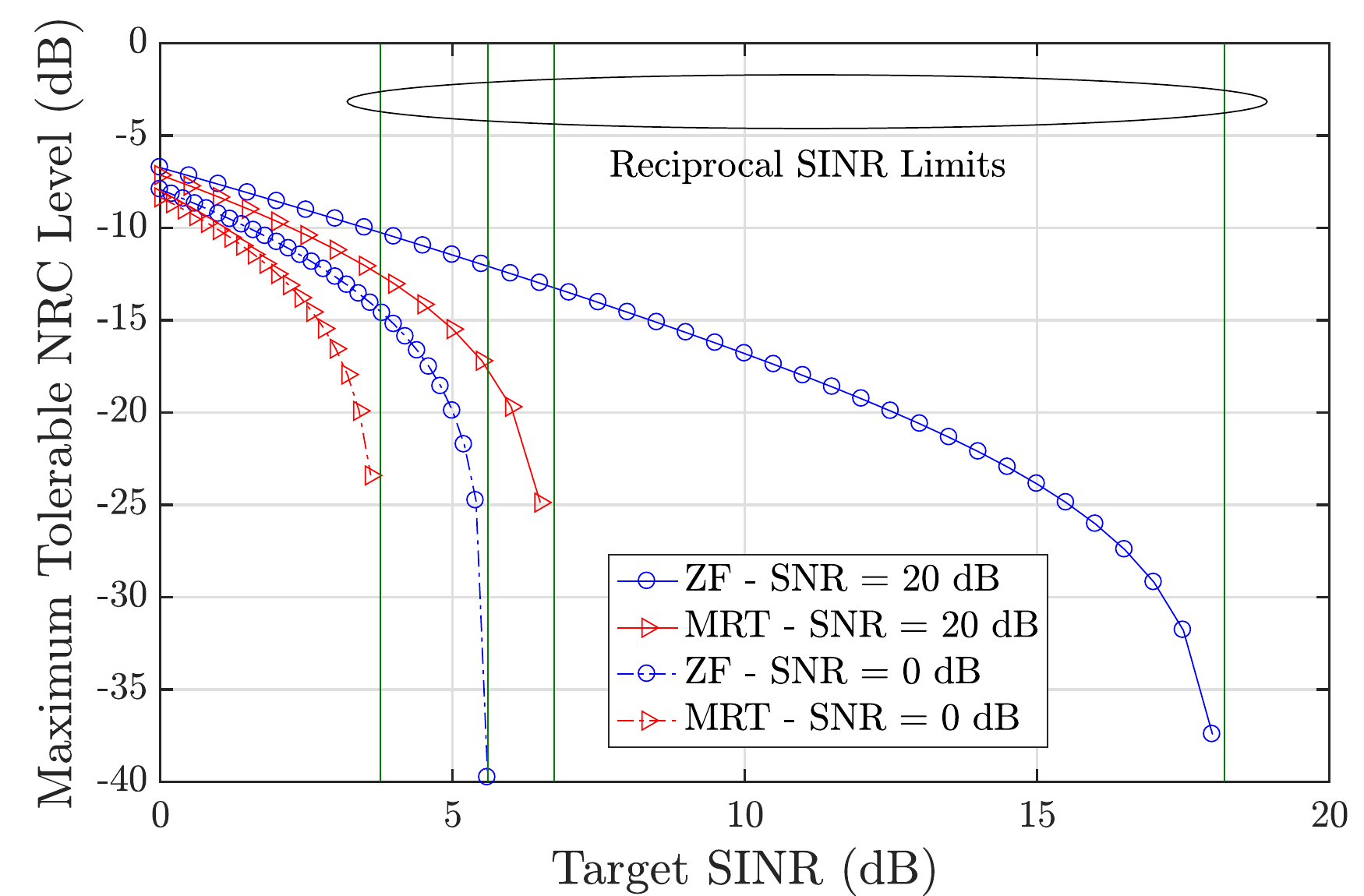}
	\caption{Maximum tolerable non-reciprocity level vs. target SINR ({$\sigma_{{a}^\prime_{d}}^2 = \sigma_{{c}^\prime_{d}}^2 = $} NRC level, while {$\sigma_{{c}^\prime_{od}}^2 = \delta_{{c}^\prime_{d}}^2 = $} NRC level $ - 10$ dB) for $N = 100$, {$M_{tot} = 20$, $K = 20$}, $\tau_u = {M_{tot}}$, $\rho_u = 0$ dB, $T = 196$.}
	\label{fig:maxNRC}
\end{figure}
In \figurename{~\ref{fig:maxNRC}}, based on the derived closed-form expressions for SINR in \eqref{ZF SINR} and \eqref{MRT SINR}, the maximum tolerable NRC level is evaluated as a function of target SINR in the UE side, for two example values of DL SNR, namely, $\rho_d = 20$ dB, $0$ dB. Based on the obtained results, in order to have SINR at UEs for example equal to $15$ dB {in ZF precoded system when $\rho_d = 20$ dB}, the maximum NRC level which can be tolerated is around {$-20$} dB. This demonstrates the value and applicability of the provided analytical results in practical system design and deployments, in for example evaluating and extracting the required NRC calibration levels such that given DL transmission performance can be achieved.

\subsection{Summary of New Findings and Future Work} \label{sec:findings}
In this subsection, we briefly summarize the novel scientific findings and concrete technical contributions of this work compared to the existing literature regarding the performance of massive MIMO systems with practical mismatched transceiver chains and antenna systems:
\begin{enumerate}
\item Based on \eqref{eq:gen performance}, for the same channel non-reciprocity levels, ZF outperforms MRT in terms of the SINR and achievable rates. However, based on derived expressions in \eqref{eq:deg}, the performance difference between the two precoding techniques starts to reduce as the level of channel non-reciprocity grows.
\item In previous literature, UE side {non-reciprocity} was assumed to have negligible effect on the total received interference \cite{ourGlobecom14}. However, this is only true when DL demodulation pilots are used to further enhance the detection at UEs. On the other hand, when UEs rely only on statistical channel properties, the UE side {non-reciprocity} has significant contribution to total received interference power. As can be inferred from derived expressions in \eqref{eq:I_ZF_NRC} and \eqref{eq:I_MRT_NRC}, for both precoding techniques, this contribution scales with ${N}$ which is a large number in the massive MIMO framework.
\item The received SINR and achievable rates of ZF and MRT precoded systems under NRC saturate at a finite and identical value asymptotically with increasing $N$. This is different from the reciprocal case where adding more antennas decreases the residual IUI and hence increases the spatial separation of UEs. This NRC-induced saturation phenomenon is due to the additional interference caused by adding more mismatched transceivers {and antenna units} with increasing $N$. 
\item Optimal number of scheduled {single-antenna UEs} under MRT is higher than that with ZF when considering moderate channel non-reciprocity levels. This is in contrast to the ideal reciprocal case where the optimal number of scheduled users is always higher for ZF precoding scheme \cite{Marzetta2}.
\end{enumerate}

In general, in addition to the channel non-reciprocity problem, pilot contamination \cite{TDDprefer} {and interference non-reciprocity \cite{Antti_NRC}} can easily be performance limiting factors, especially in multi-cell systems. Thus, joint consideration of these aspects together with NRC is an interesting research topic for our future work. Furthermore, extending the work to cover also more elaborate precoders in multi-antenna UE context, such as block-diagonalization, together with DL demodulation CSI acquisition, are interesting and important topics.

\section{Conclusion} \label{sec:conclusions}
Closed-form performance analysis of TDD-based linearly precoded massive MIMO DL system under channel non-reciprocity and imperfect CSI was carried out in this paper. The derived analytical SINR and achievable rate expressions show that in general ZF precoding scheme is more sensitive to NRC levels compared to MRT. The derived analytical expressions also show that {with inaccurate NRC calibration}, the performance gap between the two precoders decreases significantly. Moreover, in contrast to ideal reciprocal case, it was shown that the SINRs and achievable rates saturate to a finite and identical level with increasing antenna array size. Overall, the derived analytical expressions provide fundamentally useful and generic tools in dimensioning and designing practical massive MIMO systems with given performance targets, e.g., choosing the appropriate precoder based on performance-complexity trade-off, deciding the number of active antenna elements, and/or extracting the needed frequency and accuracy of adopted NRC calibration schemes.

\appendix \label{app}
In order to calculate SINR in \eqref{SINR}, we need to compute the powers of the different interference terms, namely, $z_{m}^\mathrm{SI}$ and $z_{m}^\mathrm{{ISI}}$, under ZF and MRT precoding schemes. In the continuation, the following properties and approximations are used.

\begin{itemize}
	\item \textit{Property 1}:\label{prop1}
	\begin{equation}
	\sum\limits_{l = 1}^N \sum\limits_{p = 1}^N \mathbb{E}{\left[\epsilon_{ml}\epsilon_{mp}^\mathrm{*}\right]} = \sum\limits_{l = 1}^N \mathbb{E}{\left[\epsilon_{ml}\epsilon_{ml}^\mathrm{*}\right]},
	\end{equation}
	since $\mathbb{E}{\left[\epsilon_{ml}\epsilon_{mp}^\mathrm{*}\right]} = 0$ for $l \neq p$.
	\item \textit{Property 2}:\label{prop2}
	\begin{equation}
	\sum\limits_{l = 1}^N \sum\limits_{p = 1}^N \mathbb{E}{\left[{u}_{lm}^\mathrm{ZF}{u}_{pm}^{\mathrm{ZF}^\mathrm{*}}\right]} = \sum\limits_{l = 1}^N \mathbb{E}{\left[{u}_{lm}^\mathrm{ZF}{u}_{lm}^{\mathrm{ZF}^\mathrm{*}}\right]},
	\end{equation}
	since $\mathbb{E}{\left[{u}_{lm}^\mathrm{ZF}{u}_{pm}^{\mathrm{ZF}^\mathrm{*}}\right]} = 0$ for $l \neq p$.
	\item \textit{Property 3}:\label{prop3}
	\begin{equation}
	\sum\limits_{l = 1}^N \sum\limits_{p = 1}^N \mathbb{E}{\left[\hat{h}_{ml}\hat{h}_{mp}^\mathrm{*}\right]} = \sum\limits_{l = 1}^N \mathbb{E}{\left[\hat{h}_{ml}\hat{h}_{ml}^\mathrm{*}\right]},
	\end{equation}
	since $\mathbb{E}{\left[\hat{h}_{ml}\hat{h}_{mp}^\mathrm{*}\right]} = 0$ for $l \neq p$.
	\item \textit{Approximation 1}:\label{eq:approx}
	For mathematical tractability, we employ the following approximation \cite{idealnonRecip2}
	\begin{equation}
	{u}_{li}^\mathrm{ZF} \approx \frac{\hat{{h}}_{il}^\mathrm{*}}{v^\mathrm{ZF}},
	\end{equation}
	where $v^\mathrm{ZF}$ is a constant that is chosen to satisfy $\mathbb{E}{\left[\left|{u}_{li}^{\mathrm{ZF}}\right|^2\right]} = \frac{1}{N{M_{tot}}}\mathbb{E}{\left[{\mathrm{Tr}\left({\mathbf{U}}^{{\mathrm{ZF}}^\mathrm{H}}{\mathbf{U}^{\mathrm{ZF}}}\right)}\right]}$, and hence can be expressed as
	\begin{equation}
	v^\mathrm{ZF} = \sqrt{N\left(N-{M_{tot}}\right)}\frac{\tau_u \rho_u}{\tau_u \rho_u + 1}.
	\end{equation}
	{While allowing us to derive the analytical closed-form expressions, the high accuracy of this approximation is demonstrated by the excellent match of the analytical and empirical results in Section \ref{sec:results}.}
\end{itemize}

\subsection{Interference Powers under ZF Precoding} \label{App:ZF}
Based on \eqref{eq:SI}, \eqref{eq:ZF_precoder}, and \eqref{ZF useful}, the power of the self interference can be expressed as
\begin{equation}\label{eq:z_SI_ZF}
\begin{aligned}
\mathrm{Var}\left(z_{m}^\mathrm{SI,ZF}\right) &= \mathbb{E}\left[\left|\sqrt{\rho_d}{\beta^{\mathrm{ZF}}} \sum\limits_{l \in \mathrm{UE}_k} a_{ml} \left(\hat{\mathbf{h}}_l^\mathrm{T} + {\bm{\varepsilon}}_l^\mathrm{T}\right) \mathbf{C}{{\mathbf{u}}}_{m}^{\mathrm{ZF}}s_m - \sqrt{\rho_d}{\beta^{\mathrm{ZF}}}s_m\right|^2\right]\\
&= \rho_d {\left(\beta^{\mathrm{ZF}}\right)^2} \underbrace{\mathbb{E}{\left[\left|a_{mm}^\prime s_m + \sum\limits_{\substack{l \in \mathrm{UE}_k\\l \ne m}} a_{ml} \hat{\mathbf{h}}_l^\mathrm{T} {{\mathbf{u}}}_{m}^{\mathrm{ZF}}s_m\right|^2\right]}}_{t^{\mathrm{SI,ZF}}_1} + \rho_d {\left(\beta^{\mathrm{ZF}}\right)^2} \underbrace{\mathbb{E}{\left[\left|\sum\limits_{l \in \mathrm{UE}_k} a_{ml} \hat{\mathbf{h}}_l^\mathrm{T} \mathbf{C}^\prime{{\mathbf{u}}}_{m}^{\mathrm{ZF}}s_m \right|^2\right]}}_{t^{\mathrm{SI,ZF}}_2}\\
&+ \rho_d {\left(\beta^{\mathrm{ZF}}\right)^2} \underbrace{\mathbb{E}{\left[\left|\sum\limits_{l \in \mathrm{UE}_k} a_{ml} {\bm{\varepsilon}}_l^\mathrm{T}\mathbf{C}{{\mathbf{u}}}_{m}^{\mathrm{ZF}}s_m \right|^2\right]}}_{t^{\mathrm{SI,ZF}}_3}.
\end{aligned}
\end{equation}

Next we will derive analytical expressions for the terms $t^{\mathrm{SI,ZF}}_1$, $t^{\mathrm{SI,ZF}}_2$, and $t^{\mathrm{SI,ZF}}_3$. Starting with $t^{\mathrm{SI,ZF}}_1$, we obtain
\begin{equation}\label{eq:z_SI_ZF1}
\begin{aligned}
t^{\mathrm{SI,ZF}}_1 &= \mathbb{E}\left[\left(a_{mm}^\prime s_m + \sum\limits_{\substack{l \in \mathrm{UE}_k\\l \ne m}} \sum\limits_{p = 1}^N a_{ml} \hat{{h}}_{lp} {{{u}}}_{pm}^{\mathrm{ZF}}s_m\right) \left(a_{mm}^\prime s_m + \sum\limits_{\substack{q \in \mathrm{UE}_k\\q \ne m}} \sum\limits_{r = 1}^N a_{mq} \hat{{h}}_{qr} {{{u}}}_{rm}^{\mathrm{ZF}}s_m\right)^\mathrm{*}\right]\\
&\approx \mathbb{E}{\left[\left|a_{mm}^\prime\right|^2\right]} + \frac{1}{{\left(v^\mathrm{ZF}\right)^2}}\sum\limits_{\substack{l \in \mathrm{UE}_k\\l \ne m}} \sum\limits_{p = 1}^N \mathbb{E}\left[\left|a_{ml}\right|^2\right]\mathbb{E}\left[\left|\hat{h}_{lp}\right|^2\right]\mathbb{E}\left[\left|\hat{{h}}_{mp}\right|^2\right]\\
&= \sigma_{{a}^\prime_{mm}}^2 + \frac{1}{N-{M_{tot}}}\left(\mathrm{Tr}\left(\mathbf{R}_{\mathbf{a}^\prime_{m}}\right) - \sigma_{{a}^\prime_{mm}}^2\right).
\end{aligned}
\end{equation}
In obtaining the expression on the second line, we used Approximation 1 and Property 3.

Following that, $t^{\mathrm{SI,ZF}}_2$ can be expressed as
\begin{equation}\label{eq:z_SI_ZF2}
\begin{aligned}
t^{\mathrm{SI,ZF}}_2 &\approx \frac{1}{{\left(v^\mathrm{ZF}\right)^2}} \sum\limits_{\substack{l \in \mathrm{UE}_k}} \sum\limits_{\substack{q \in \mathrm{UE}_k}} \sum\limits_{p = 1}^N \sum\limits_{r = 1}^N \sum\limits_{i = 1}^N \sum\limits_{j = 1}^N \mathbb{E}\left[a_{ml}a_{mq}^\mathrm{*}\right] \mathbb{E}\left[\hat{h}_{lp}\hat{{h}}_{mr}^\mathrm{*}\hat{h}_{qi}^\mathrm{*}\hat{{h}}_{mj}\right] \mathbb{E}{\left[c_{pr}^\prime c_{ij}^{\prime^\mathrm{*}}\right]}\\
&= \frac{1}{{\left(v^\mathrm{ZF}\right)^2}} \underbrace{\sum\limits_{p = 1}^N \sum\limits_{r = 1}^N \sum\limits_{i = 1}^N \sum\limits_{j = 1}^N \mathbb{E}\left[\left|a_{mm}\right|^2\right] \mathbb{E}\left[\hat{h}_{mp}\hat{h}_{mi}^\mathrm{*}\hat{{h}}_{mr}^\mathrm{*}\hat{{h}}_{mj}\right] \mathbb{E}{\left[c_{pr}^\prime c_{ij}^{\prime^\mathrm{*}}\right]}}_{t^{\mathrm{SI,ZF}}_{21}}\\
&+ \frac{1}{{\left(v^\mathrm{ZF}\right)^2}} \sum\limits_{\substack{l \in \mathrm{UE}_k\\l \neq m}} \sum\limits_{p = 1}^N \sum\limits_{r = 1}^N \mathbb{E}\left[\left|a_{ml}\right|^2\right] \mathbb{E}\left[\left|\hat{h}_{lp}\right|^2\right] \mathbb{E}\left[\left|\hat{{h}}_{mr}\right|^2\right] \mathbb{E}{\left[\left|c_{pr}^\prime\right|^2\right]}\\
&= \frac{1}{{\left(v^\mathrm{ZF}\right)^2}} \left(t^{\mathrm{SI,ZF}}_{21} + \left(\mathrm{Tr}\left(\mathbf{R}_{\mathbf{a}^\prime_{m}}\right) - \sigma_{{a}^\prime_{mm}}^2\right) \left(\frac{\tau_u \rho_u}{\tau_u \rho_u + 1}\right)^2 \left(\mathrm{Tr}\left(\mathbf{R}_{\mathrm{c}^\prime_{d}}\right) + \mathrm{Tr}\left(\mathbf{R}_{\mathrm{c}^\prime_{od}}\right)\right) \right).
\end{aligned}
\end{equation}
In above, we used Approximation 1 when obtaining the expression on the first line, whereas the expression on the second and the third lines are obtained using Property 3. In the next step, ${t^{\mathrm{SI,ZF}}_{21}}$ is expressed as
\begin{equation} \label{eq:app_t21_SI_ZF}
\begin{aligned}
t^{\mathrm{SI,ZF}}_{21} &= \mathbb{E}\left[\left|a_{mm}\right|^2\right] \sum\limits_{p = 1}^N \sum\limits_{\substack{r = 1\\r \neq p}}^N \mathbb{E}\left[\left|\hat{h}_{mp}\right|^2\right] \mathbb{E}\left[\left|\hat{{h}}_{mr}\right|^2\right] \mathbb{E}{\left[\left|c_{pr}^\prime\right|^2\right]}\\
&+ \mathbb{E}\left[\left|a_{mm}\right|^2\right] \sum\limits_{p = 1}^N \sum\limits_{\substack{j = 1\\j \neq p}}^N \mathbb{E}\left[\left|\hat{h}_{mp}\right|^2\right] \mathbb{E}\left[\left|\hat{{h}}_{mj}\right|^2\right] \mathbb{E}{\left[c_{pp}^\prime c_{jj}^{\prime^\mathrm{*}}\right]} + \mathbb{E}\left[\left|a_{mm}\right|^2\right] \sum\limits_{p = 1}^N \mathbb{E}\left[\left|\hat{h}_{mp}\right|^4\right] \mathbb{E}{\left[\left|c_{pp}^\prime\right|^2\right]}\\
&= \left(1 + \sigma_{{a}^\prime_{mm}}^2\right) \left(\frac{\tau_u \rho_u}{\tau_u \rho_u + 1}\right)^2 \left(\mathrm{Tr}\left(\mathbf{R}_{\mathrm{c}^\prime_{od}}\right) + \mathrm{Sum}\left(\mathbf{R}_{\mathrm{c}^\prime_{d}}\right) + \mathrm{Tr}\left(\mathbf{R}_{\mathrm{c}^\prime_{d}}\right)\right),
\end{aligned}
\end{equation}
where Property 3 is used in obtaining the expression in the first two lines.

Substituting \eqref{eq:app_t21_SI_ZF} in \eqref{eq:z_SI_ZF2}, we have
\begin{equation}
t^{\mathrm{SI,ZF}}_2 \approx \frac{1}{N\left(N-{M_{tot}}\right)} \left(\left(1 + \sigma_{{a}^\prime_{mm}}^2\right) \mathrm{Sum}\left(\mathbf{R}_{\mathrm{c}^\prime_{d}}\right) + \left(1 + \mathrm{Tr}\left(\mathbf{R}_{\mathbf{a}^\prime_{m}}\right)\right) \left(\mathrm{Tr}\left(\mathbf{R}_{\mathrm{c}^\prime_{d}}\right) + \mathrm{Tr}\left(\mathbf{R}_{\mathrm{c}^\prime_{od}}\right)\right)\right).
\end{equation}

Finally, the term $t^{\mathrm{SI,ZF}}_3$ can be expressed as
\begin{equation}\label{eq:z_SI_ZF3}
\begin{aligned}
t^{\mathrm{SI,ZF}}_3 &= \sum\limits_{\substack{l \in \mathrm{UE}_k}} \sum\limits_{p = 1}^N \sum\limits_{r = 1}^N \mathbb{E}\left[\left|a_{ml}\right|^2\right] \mathbb{E}\left[\left|\epsilon_{lp}\right|^2\right] \mathbb{E}{\left[\left|c_{pr}\right|^2\right]} \mathbb{E}\left[\left|{u}_{rm}^\mathrm{ZF}\right|^2\right]\\
&= \frac{1 + \mathrm{Tr}\left(\mathbf{R}_{\mathbf{a}^\prime_{m}}\right)}{N{M_{tot}}\left(\beta^{\mathrm{ZF}}\right)^2\left(\tau_u \rho_u + 1\right)} \left(N + \mathrm{Tr}\left(\mathbf{R}_{\mathrm{c}^\prime_{d}}\right) + \mathrm{Tr}\left(\mathbf{R}_{\mathrm{c}^\prime_{od}}\right)\right).
\end{aligned}
\end{equation}
In obtaining the expression on the first line, we used Property 1 and Property 2.

Similarly, based on \eqref{eq:SI}, the power of the {ISI} under ZF precoding scheme can be written as
\begin{equation} \label{eq:z_IUI_ZF}
\begin{aligned}
\mathrm{Var}&\left(z_{m}^\mathrm{{ISI},ZF}\right) = \mathbb{E}{\left[\left|\sqrt{\rho_d}{\beta^{\mathrm{ZF}}} \sum\limits_{\substack{i = 1\\i \ne m}}^{M_{tot}} \sum\limits_{l \in \mathrm{UE}_k} a_{ml} \left(\hat{\mathbf{h}}_l^\mathrm{T} + {\bm{\varepsilon}}_l^\mathrm{T}\right) \mathbf{C}{{\mathbf{u}}}_{i}^{\mathrm{ZF}}s_i\right|^2\right]}\\
&= \rho_d {\left(\beta^{\mathrm{ZF}}\right)^2} \underbrace{\mathbb{E}{\left[\left|\sum\limits_{\substack{i \in \mathrm{UE}_k\\i \ne m}} a_{mi} s_i\right|^2\right]}}_{t^{\mathrm{{ISI},ZF}}_1} + \rho_d {\left(\beta^{\mathrm{ZF}}\right)^2} \underbrace{\mathbb{E}{\left[\left|\sum\limits_{\substack{i = 1\\i \ne m}}^{M_{tot}} \sum\limits_{l \in \mathrm{UE}_k} a_{ml} \hat{\mathbf{h}}_l^\mathrm{T} \mathbf{C}^\prime{{\mathbf{u}}}_{i}^{\mathrm{ZF}}s_i \right|^2\right]}}_{t^{\mathrm{{ISI},ZF}}_2}\\
&+ \rho_d {\left(\beta^{\mathrm{ZF}}\right)^2} \underbrace{\mathbb{E}{\left[\left|\sum\limits_{\substack{i = 1\\i \ne m}}^{M_{tot}} \sum\limits_{l \in \mathrm{UE}_k} a_{ml} \bm{\varepsilon}_l^\mathrm{T} \mathbf{C}{{\mathbf{u}}}_{i}^{\mathrm{ZF}}s_i \right|^2\right]}}_{t^{\mathrm{{ISI},ZF}}_3}.
\end{aligned}
\end{equation}

Next, we will derive analytical expressions for the terms $t^{\mathrm{{ISI},ZF}}_1$, $t^{\mathrm{{ISI},ZF}}_2$ and $t^{\mathrm{{ISI},ZF}}_3$. Starting with $t^{\mathrm{{ISI},ZF}}_1$, we obtain
\begin{equation}\label{eq:z_IUI_ZF1}
t^{\mathrm{{ISI},ZF}}_1 = \sum\limits_{\substack{i \in \mathrm{UE}_k\\i \ne m}} \mathbb{E}{\left[\left|a_{mi}\right|^2\right]} \mathbb{E}{\left[\left|s_i\right|^2\right]} = \mathrm{Tr}\left(\mathbf{R}_{\mathbf{a}^\prime_{m}}\right) - \sigma_{{a}^\prime_{mm}}^2.
\end{equation}

Following that, $t^{\mathrm{{ISI},ZF}}_2$ can be expressed as
\begin{equation}\label{eq:z_IUI_ZF2}
\begin{aligned}
t^{\mathrm{{ISI},ZF}}_2 &\approx \frac{1}{{\left(v^\mathrm{ZF}\right)^2}} \sum\limits_{\substack{i = 1\\i \ne m}}^{M_{tot}} \sum\limits_{\substack{l \in \mathrm{UE}_k}} \sum\limits_{p = 1}^N \sum\limits_{r = 1}^N \sum\limits_{o = 1}^N \sum\limits_{w = 1}^N \mathbb{E}\left[\left|a_{ml}\right|^2\right] \mathbb{E}\left[\hat{h}_{lp}\hat{{h}}_{ir}^\mathrm{*}\hat{h}_{lo}^\mathrm{*}\hat{{h}}_{iw}\right] \mathbb{E}{\left[c_{pr}^\prime c_{ow}^{\prime^\mathrm{*}}\right]}\\
&= \frac{1}{{\left(v^\mathrm{ZF}\right)^2}} \sum\limits_{\substack{i \in \mathrm{UE}_k\\i \ne m}} \mathbb{E}\left[\left|a_{mi}\right|^2\right] \frac{t^{\mathrm{SI,ZF}}_{21}}{\mathbb{E}\left[\left|a_{mm}\right|^2\right]}\\
&+ \frac{1}{{\left(v^\mathrm{ZF}\right)^2}} \left(\sum\limits_{\substack{i = 1\\i \notin \mathrm{UE}_k}}^{M_{tot}} \left(1 + \mathrm{Tr}\left(\mathbf{R}_{\mathbf{a}^\prime_{m}}\right)\right) + \sum\limits_{\substack{i \in \mathrm{UE}_k\\i \ne m}} \left({1+}\mathrm{Tr}\left(\mathbf{R}_{\mathbf{a}^\prime_{m}}\right) - \sigma_{a_{mi}}^2\right)\right) \left(\frac{\tau_u \rho_u}{\tau_u \rho_u + 1}\right)^2 \left(\mathrm{Tr}\left(\mathbf{R}_{\mathrm{c}^\prime_{d}}\right) + \mathrm{Tr}\left(\mathbf{R}_{\mathrm{c}^\prime_{od}}\right)\right)\\
&= \frac{1}{N\left(N-{M_{tot}}\right)} \left(\left(\left(1+\sigma_{{a}^\prime_{mm}}^2\right) + \left({M_{tot}}-2\right) \left(1+\mathrm{Tr}\left(\mathbf{R}_{\mathbf{a}^\prime_{m}}\right)\right) \right) \left(\mathrm{Tr}\left(\mathbf{R}_{\mathrm{c}^\prime_{d}}\right) + \mathrm{Tr}\left(\mathbf{R}_{\mathrm{c}^\prime_{od}}\right)\right) \right.\\
&+ \left. \left(\mathrm{Tr}\left(\mathbf{R}_{\mathbf{a}^\prime_{m}}\right) - \sigma_{{a}^\prime_{mm}}^2\right) \mathrm{Sum}\left(\mathbf{R}_{\mathrm{c}^\prime_{d}}\right)\right).
\end{aligned}
\end{equation}
In above, we used the Approximation 1 in obtaining the expression on the first line and Property 3 in obtaining the expressions on the first three lines.

Then, $t^{\mathrm{{ISI},ZF}}_3$ can be expressed, similar to $t^{\mathrm{SI,ZF}}_3$, as
\begin{equation} \label{eq:z_IUI_ZF3}
t^{\mathrm{{ISI},ZF}}_3 = \left({M_{tot}}-1\right)\frac{1 + \mathrm{Tr}\left(\mathbf{R}_{\mathbf{a}^\prime_{m}}\right)}{N{M_{tot}}\left(\beta^{\mathrm{ZF}}\right)^2\left(\tau_u \rho_u + 1\right)} \left(N + \mathrm{Tr}\left(\mathbf{R}_{\mathrm{c}^\prime_{d}}\right) + \mathrm{Tr}\left(\mathbf{R}_{\mathrm{c}^\prime_{od}}\right)\right).
\end{equation}
Here, we used Property 1 and Property 2.

The total interference power can be obtained by summing all the calculated interference terms. Then, it is straightforward to re-arrange the terms and express the total interference power as ${{I}_{\mathrm{RC}}^{\mathrm{ZF}}} + {I}_{\mathrm{NRC},m}^{\mathrm{ZF}}$ after which we reach the SINR expression presented in \eqref{ZF SINR}.

\subsection{Interference Powers under MRT Precoding} \label{App:MRT}
Based on \eqref{eq:SI}, \eqref{eq:MRT_precoder}, and \eqref{MRT useful}, the power of self interference under MRT precoding scheme can be expressed as
\begin{equation}\label{eq:z_SI_MRT}
\begin{aligned}
\mathrm{Var}\left(z_{m}^\mathrm{SI,MRT}\right) &= \mathbb{E}\left[\left|\sqrt{\rho_d}{\beta^{\mathrm{MRT}}} \sum\limits_{l \in \mathrm{UE}_k} a_{ml} \left(\hat{\mathbf{h}}_l^\mathrm{T} + {\bm{\varepsilon}}_l^\mathrm{T}\right) \mathbf{C}{{\mathbf{u}}}_{m}^{\mathrm{MRT}}s_m - \sqrt{\rho_d}{\beta^{\mathrm{MRT}}}\mathbb{E}\left[\mathbf{h}_m^\mathrm{T}{{\mathbf{u}}}_{m}^{\mathrm{MRT}}\right]s_m\right|^2\right]\\
&= \rho_d {\left(\beta^{\mathrm{MRT}}\right)^2} \underbrace{\mathbb{E}\left[\left|a_{mm} \hat{\mathbf{h}}_m^\mathrm{T}\hat{\mathbf{h}}_m^\mathrm{*} s_m - \mathbb{E}\left[\hat{\mathbf{h}}_m^\mathrm{T}\hat{\mathbf{h}}_m^\mathrm{*}\right]s_m + \sum\limits_{\substack{l \in \mathrm{UE}_k\\l \ne m}} a_{ml} \hat{\mathbf{h}}_l^\mathrm{T} \hat{\mathbf{h}}_m^\mathrm{*}s_m\right|^2\right]}_{t^{\mathrm{SI,MRT}}_1} \\ 
&+ \rho_d {\left(\beta^{\mathrm{MRT}}\right)^2} \underbrace{\mathbb{E}{\left[\left|\sum\limits_{l \in \mathrm{UE}_k} a_{ml} \hat{\mathbf{h}}_l^\mathrm{T} \mathbf{C}^\prime\hat{\mathbf{h}}_m^\mathrm{*}s_m \right|^2\right]}}_{t^{\mathrm{SI,MRT}}_2} + \rho_d {\left(\beta^{\mathrm{MRT}}\right)^2} \underbrace{\mathbb{E}{\left[\left|\sum\limits_{l \in \mathrm{UE}_k} a_{ml} {\bm{\varepsilon}}_l^\mathrm{T}\mathbf{C}\hat{\mathbf{h}}_m^\mathrm{*}s_m \right|^2\right]}}_{t^{\mathrm{SI,MRT}}_3}.
\end{aligned}
\end{equation}

Next we derive analytical expressions for the terms $t^{\mathrm{SI,MRT}}_1, t^{\mathrm{SI,MRT}}_2$ and $t^{\mathrm{SI,MRT}}_3$. Starting with $t^{\mathrm{SI,MRT}}_1$, we get
\begin{equation}\label{eq:z_SI_MRT1}
t^{\mathrm{SI,MRT}}_1 = \underbrace{\mathbb{E}{\left[\left|a_{mm} \hat{\mathbf{h}}_m^\mathrm{T}\hat{\mathbf{h}}_m^\mathrm{*} s_m - \mathbb{E}\left[\hat{\mathbf{h}}_m^\mathrm{T}\hat{\mathbf{h}}_m^\mathrm{*}\right]s_m \right|^2\right]}}_{t^{\mathrm{SI,MRT}}_{11}} + \underbrace{\mathbb{E}{\left[\left|\sum\limits_{\substack{l \in \mathrm{UE}_k\\l \ne m}} a_{ml} \hat{\mathbf{h}}_l^\mathrm{T} \hat{\mathbf{h}}_m^\mathrm{*}s_m\right|^2\right]}}_{t^{\mathrm{SI,MRT}}_{12}}.
\end{equation}

Following that, $t^{\mathrm{SI,MRT}}_{11}$ can be expressed as
\begin{equation}\label{eq:z_SI_MRT11}
\begin{aligned}
t^{\mathrm{SI,MRT}}_{11} &= \mathbb{E}{\left[\left|a_{mm} \hat{\mathbf{h}}_m^\mathrm{T}\hat{\mathbf{h}}_m^\mathrm{*} s_m \right|^2\right]} - \mathbb{E}{\left[\left|\mathbb{E}\left[\hat{\mathbf{h}}_m^\mathrm{T}\hat{\mathbf{h}}_m^\mathrm{*}\right]s_m \right|^2\right]}\\
&= \mathbb{E}{\left[\left|a_{mm}\right|^2\right]} \left(\sum\limits_{p = 1}^N \sum\limits_{\substack{r = 1\\r \ne p}}^N \mathbb{E}{\left[\left|\hat{h}_{mp}\right|^2\right]}\mathbb{E}{\left[\left|\hat{h}_{mr}\right|^2\right]} + \sum\limits_{p = 1}^N \mathbb{E}{\left[\left|\hat{h}_{mp}\right|^4\right]}\right) - N^2 \left(\frac{\tau_u \rho_u}{\tau_u \rho_u + 1}\right)^2\\
&= N\left(1 + \sigma_{{a}^\prime_{mm}}^2 \left(N+1\right)\right) \left(\frac{\tau_u \rho_u}{\tau_u \rho_u + 1}\right)^2.
\end{aligned}
\end{equation}

Next we express $t^{\mathrm{SI,MRT}}_{12}$ as
\begin{equation}\label{eq:z_SI_MRT12}
t^{\mathrm{SI,MRT}}_{12} = \sum\limits_{\substack{l \in \mathrm{UE}_k\\l \ne m}} \sum\limits_{p = 1}^N \mathbb{E}{\left[\left|a_{ml}\right|^2\right]} \mathbb{E}{\left[\left|\hat{h}_{lp}\right|^2\right]}\mathbb{E}{\left[\left|\hat{h}_{mp}\right|^2\right]} = N \left(\mathrm{Tr}\left(\mathbf{R}_{\mathbf{a}^\prime_{m}}\right) - \sigma_{{a}^\prime_{mm}}^2\right) \left(\frac{\tau_u \rho_u}{\tau_u \rho_u + 1}\right)^2,
\end{equation}
where Property 3 is used in obtaining the expression on the first line.

Substituting \eqref{eq:z_SI_MRT11} and \eqref{eq:z_SI_MRT12} in \eqref{eq:z_SI_MRT1}, we have
\begin{equation}
t^{\mathrm{SI,MRT}}_1 = N \left(1 + N\sigma_{{a}^\prime_{mm}}^2 + \mathrm{Tr}\left(\mathbf{R}_{\mathbf{a}^\prime_{m}}\right)\right) \left(\frac{\tau_u \rho_u}{\tau_u \rho_u + 1}\right)^2.
\end{equation}

Then, we can express $t^{\mathrm{SI,MRT}}_2$, similar to $t^{\mathrm{SI,ZF}}_2$, as
\begin{equation}\label{eq:z_SI_MRT2}
t^{\mathrm{SI,MRT}}_2 = \left(\frac{\tau_u \rho_u}{\tau_u \rho_u + 1}\right)^2 \left(\left(1 + \sigma_{{a}^\prime_{mm}}^2\right) \mathrm{Sum}\left(\mathbf{R}_{\mathrm{c}^\prime_{d}}\right) + \left(1 + \mathrm{Tr}\left(\mathbf{R}_{\mathbf{a}^\prime_{m}}\right)\right) \left(\mathrm{Tr}\left(\mathbf{R}_{\mathrm{c}^\prime_{d}}\right) + \mathrm{Tr}\left(\mathbf{R}_{\mathrm{c}^\prime_{od}}\right)\right)\right).
\end{equation}
In obtaining the final expression, we used Property 3.

Following that, $t^{\mathrm{SI,MRT}}_3$ can be expressed, similar to $t^{\mathrm{SI,ZF}}_3$, as
\begin{equation}\label{eq:z_SI_MRT3}
t^{\mathrm{SI,MRT}}_3 = \frac{\tau_u \rho_u}{\left(\tau_u \rho_u + 1\right)^2} \left(1 + \mathrm{Tr}\left(\mathbf{R}_{\mathbf{a}^\prime_{m}}\right)\right) \left(N + \mathrm{Tr}\left(\mathbf{R}_{\mathrm{c}^\prime_{d}}\right) + \mathrm{Tr}\left(\mathbf{R}_{\mathrm{c}^\prime_{od}}\right)\right).
\end{equation}
In obtaining the final expression, we used Property 1 and Property 3.

Then, based on \eqref{eq:SI}, the power of {ISI} under MRT precoding scheme can be written as
\begin{equation} \label{eq:z_IUI_MRT}
\begin{aligned}
\mathrm{Var}\left(z_{m}^\mathrm{{ISI},MRT}\right) &= \mathbb{E}{\left[\left|\sqrt{\rho_d}{\beta^{\mathrm{MRT}}} \sum\limits_{\substack{i = 1\\i \ne m}}^{M_{tot}} \sum\limits_{l \in \mathrm{UE}_k} a_{ml} \left(\hat{\mathbf{h}}_l^\mathrm{T} + {\bm{\varepsilon}}_l^\mathrm{T}\right) \mathbf{C}{{\mathbf{u}}}_{i}^{\mathrm{MRT}}s_i\right|^2\right]}\\
&= \rho_d {\left(\beta^{\mathrm{MRT}}\right)^2} \left(\underbrace{\mathbb{E}{\left[\left|\sum\limits_{\substack{i = 1\\i \ne m}}^{M_{tot}} \sum\limits_{l \in \mathrm{UE}_k} a_{ml} \hat{\mathbf{h}}_l^\mathrm{T} \mathbf{C}\hat{\mathbf{h}}_i^\mathrm{*}s_i \right|^2\right]}}_{t^{\mathrm{{ISI},MRT}}_1} + \underbrace{\mathbb{E}{\left[\left|\sum\limits_{\substack{i = 1\\i \ne m}}^{M_{tot}} \sum\limits_{l \in \mathrm{UE}_k} a_{ml} \bm{\varepsilon}_l^\mathrm{T} \mathbf{C}\hat{\mathbf{h}}_i^\mathrm{*}s_i \right|^2\right]}}_{t^{\mathrm{{ISI},MRT}}_2}\right).
\end{aligned}
\end{equation}

Next, we will derive analytical expressions for the terms $t^{\mathrm{{ISI},MRT}}_1$ and $t^{\mathrm{{ISI},MRT}}_2$. Starting with $t^{\mathrm{{ISI},MRT}}_1$, similar to $t^{\mathrm{{ISI},ZF}}_2$, we get
\begin{equation}\label{eq:z_IUI_MRT1}
\begin{aligned}
t^{\mathrm{{ISI},MRT}}_1 &= \left(\frac{\tau_u \rho_u}{\tau_u \rho_u + 1}\right)^2 \left(\left(\left({M_{tot}}-2\right) \left(1+\mathrm{Tr}\left(\mathbf{R}_{\mathbf{a}^\prime_{m}}\right)\right) + \left(1+\sigma_{{a}^\prime_{mm}}^2\right) \right) \left(N + \mathrm{Tr}\left(\mathbf{R}_{\mathrm{c}^\prime_{d}}\right) + \mathrm{Tr}\left(\mathbf{R}_{\mathrm{c}^\prime_{od}}\right)\right)\right.\\
&+ \left. \left(\mathrm{Tr}\left(\mathbf{R}_{\mathbf{a}^\prime_{m}}\right) - \sigma_{{a}^\prime_{mm}}^2\right) \left(N^2 + \mathrm{Sum}\left(\mathbf{R}_{\mathrm{c}^\prime_{d}}\right)\right)\right).
\end{aligned}
\end{equation}
In obtaining the final expression, we used Property 3.

Following that, $t^{\mathrm{{ISI},MRT}}_2$ can be expressed, similar to $t^{\mathrm{{ISI},ZF}}_3$, as 
\begin{equation}\label{eq:z_IUI_MRT2}
t^{\mathrm{{ISI},MRT}}_2 = \left({M_{tot}}-1\right)\frac{\tau_u \rho_u}{\left(\tau_u \rho_u + 1\right)^2} \left(1 + \mathrm{Tr}\left(\mathbf{R}_{\mathbf{a}^\prime_{m}}\right)\right) \left(N + \mathrm{Tr}\left(\mathbf{R}_{\mathrm{c}^\prime_{d}}\right) + \mathrm{Tr}\left(\mathbf{R}_{\mathrm{c}^\prime_{od}}\right)\right).
\end{equation}
In obtaining the final expression, we used Property 1 and Property 3.

Finally, the total interference power is obtained by summing all the calculated interference terms. Then, it is straightforward to re-arrange the terms and express the total interference power as ${{I}_{\mathrm{RC}}^{\mathrm{MRT}}} + {I}_{\mathrm{NRC},m}^{\mathrm{MRT}}$ after which we reach the SINR expression presented in \eqref{MRT SINR}.


\ifCLASSOPTIONcaptionsoff
  \newpage
\fi



\bibliographystyle{IEEEtranTCOM}
\bibliography{IEEEabrv,refs}
%
%
%
%

%
\vspace*{-2\baselineskip}
\begin{IEEEbiography}[{\includegraphics[width=1in,height=1.25in,clip,keepaspectratio]{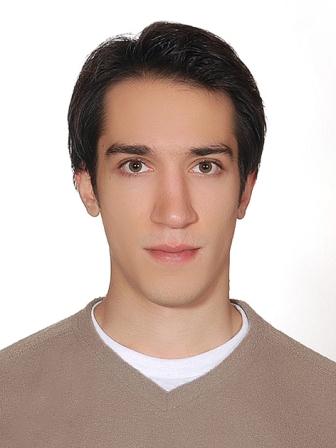}}]{Orod Raeesi}
	received his M.Sc. degree (with distinction) from Tampere University of Technology (TUT), Tampere, Finland, in 2011, and is currently pursuing the Ph.D. degree at TUT. Currently, he is working as a communications system specialist with Nokia Mobile Networks, Espoo, Finland. His research interests include IEEE~802.11 MAC and PHY layer challenges, massive MIMO systems, TDD channel non-reciprocity, ultra-reliable low latency communications, system-level simulations, and 5G mobile radio networks.
\end{IEEEbiography}
\begin{IEEEbiography}[{\includegraphics[width=1in,height=1.25in,clip,keepaspectratio]{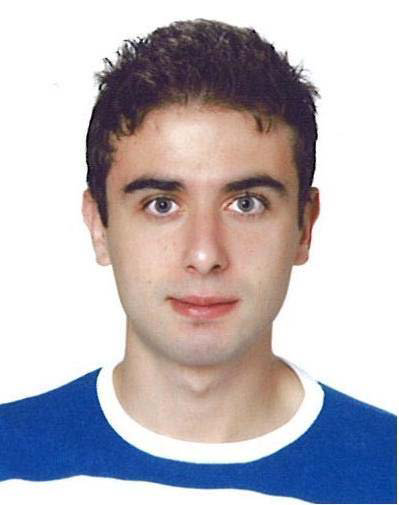}}]{Ahmet Gokceoglu}
	received M.Sc. (2010) and Ph.D. Degrees (2014) from the Department of Electronics and Communications Engineering, Tampere University of Technology, Finland, where he also held a postdoctoral researcher position (2014-2016) together with a visiting researcher position at Link\"{o}ping university (2015). His research interests are information theory, modeling and performance analysis of MIMO-OFDM systems, advanced transmitter and receiver signal processing techniques and algorithm design for MAC and physical layer. He is currently working as an algorithm specialist with Huawei, Stockholm, Sweden.
\end{IEEEbiography}
\begin{IEEEbiography}[{\includegraphics[width=1in,height=1.25in,clip,keepaspectratio]{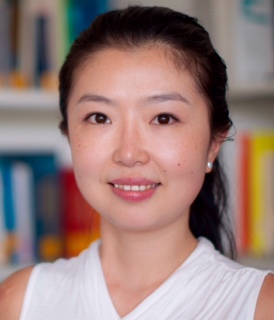}}]{Yaning Zou}
	received her Bachelor Degree in communications engineering from University Electronic Science and Technology of China (UESTC), Chengdu, China, in 2002, and the M.Sc. and Ph.D. Degrees in electrical engineering from Tampere University of Technology (TUT), Tampere, Finland, in 2005 and 2009, respectively. Then she worked as an Academy postdoctoral researcher at TUT, Finland, until August 2015. In September 2015, she joined the Vodafone Chair Mobile Communication Systems Technical University of Dresden, Germany, where she currently works as a research manager. She has published over 30 articles in international peer-reviewed journals and conferences, as well as one book chapter. Her general research interests are in physical layer design and RF implementation challenges of 5G and beyond communication system as well as related ICT Policies.
\end{IEEEbiography}
\begin{IEEEbiography}[{\includegraphics[width=1in,height=1.25in,clip,keepaspectratio]{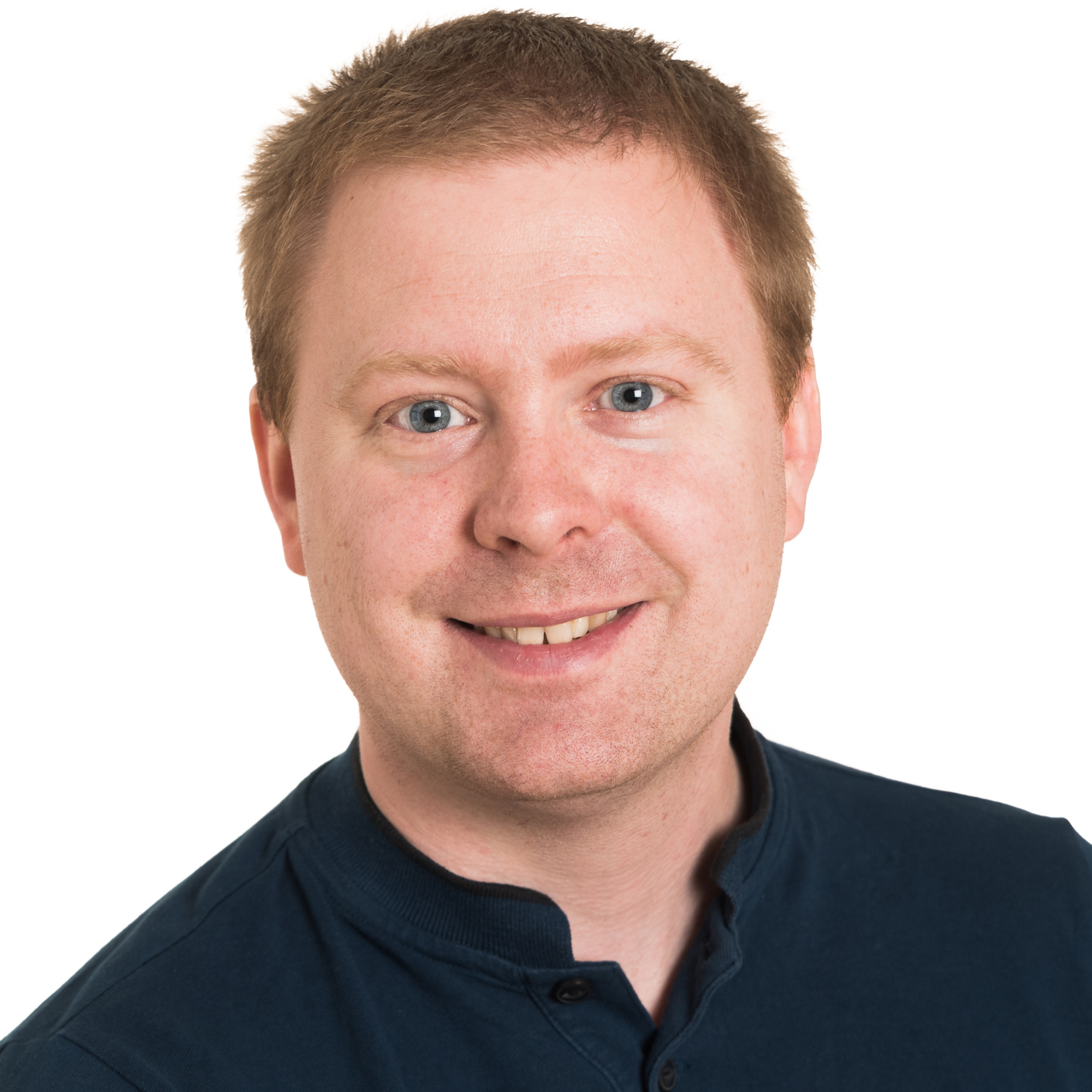}}]{Emil Bj\"{o}rnson}
	(S'07--M'12--SM'17) received the M.S. degree in Engineering Mathematics from Lund University, Sweden, in 2007. He received the Ph.D. degree in Telecommunications from KTH Royal Institute of Technology, Sweden, in 2011. From 2012 to mid 2014, he was a joint postdoc at the Alcatel-Lucent Chair on Flexible Radio, SUPELEC, France, and at KTH. He joined Link\"{o}ping University, Sweden, in 2014 and is currently Associate Professor and Docent at the Division of Communication Systems.
	He performs research on multi-antenna communications, Massive MIMO, radio resource allocation, energy-efficient communications, and network design. He is on the editorial board of the \textsc{IEEE Transactions on Communications} and the \textsc{IEEE Transactions on Green Communications and Networking}. He is the first author of the textbooks \emph{Massive MIMO Networks: Spectral, Energy, and Hardware Efficiency} (2017) and \emph{Optimal Resource Allocation in Coordinated Multi-Cell Systems} (2013). He is dedicated to reproducible research and has made a large amount of simulation code publicly available.
	Dr. Bj\"{o}rnson has performed MIMO research for more than ten years and has filed more than ten related patent applications. He received the 2016 Best PhD Award from EURASIP, the 2015 Ingvar Carlsson Award, and the 2014 Outstanding Young Researcher Award from IEEE ComSoc EMEA. He has co-authored papers that received best paper awards at WCSP 2017, IEEE ICC 2015, IEEE WCNC 2014, IEEE SAM 2014, IEEE CAMSAP 2011, and WCSP 2009.
\end{IEEEbiography}
\begin{IEEEbiography}[{\includegraphics[width=1in,height=1.25in,clip,keepaspectratio]{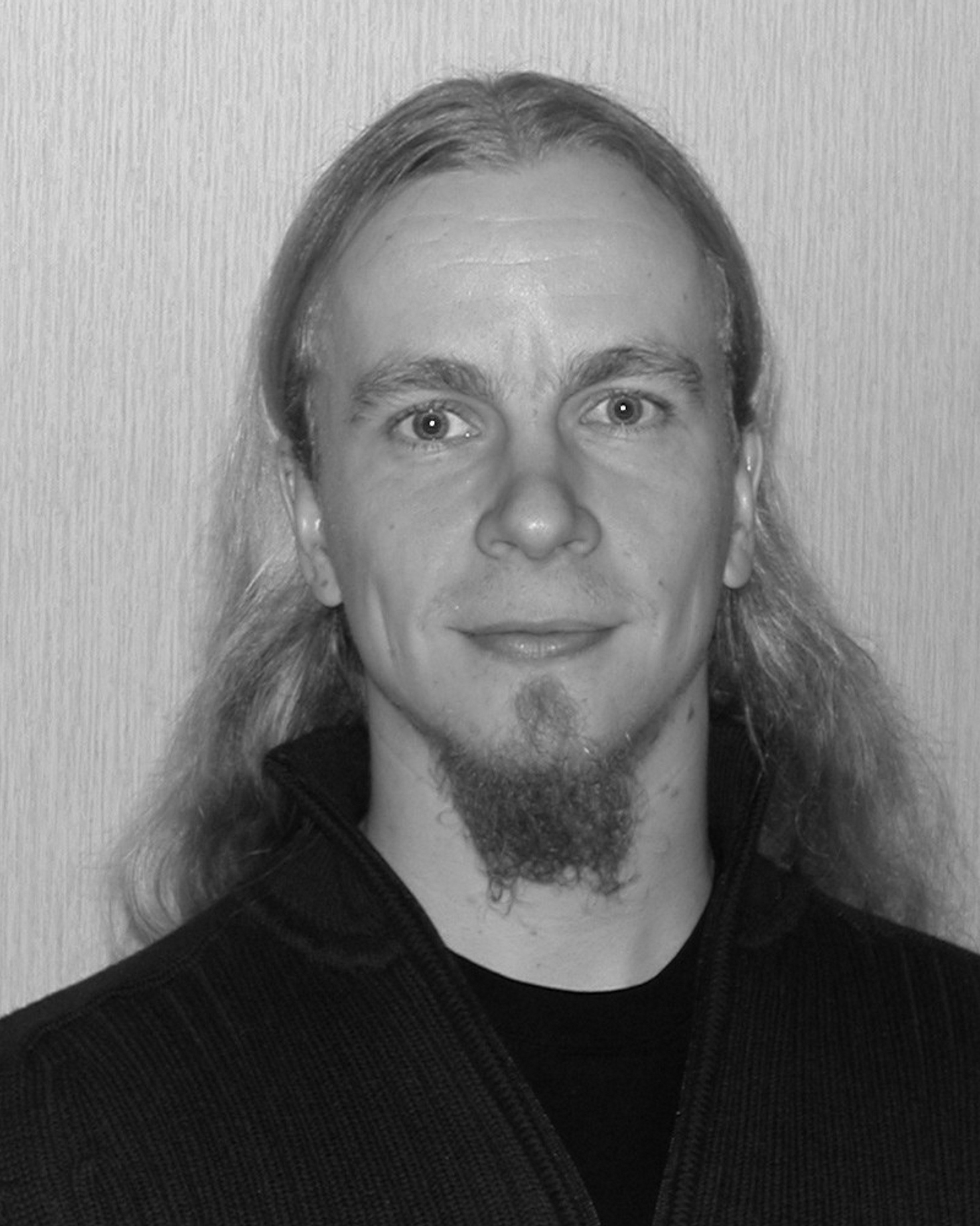}}]{Mikko Valkama}
	 (S'00--M'01--SM'15) was born in Pirkkala, Finland, on November 27, 1975. He received the M.Sc. and Ph.D. Degrees (both with honors) in electrical engineering (EE) from Tampere University of Technology (TUT), Finland, in 2000 and 2001, respectively. In 2002, he received the Best Ph.D. Thesis -award by the Finnish Academy of Science and Letters for his dissertation entitled ``Advanced I/Q signal processing for wideband receivers: Models and algorithms''. In 2003, he was working as a visiting post-doc research fellow with the Communications Systems and Signal Processing Institute at SDSU, San Diego, CA. Currently, he is a Full Professor and Laboratory Head at the Laboratory of Electronics and Communications Engineering at TUT, Finland. His general research interests include radio communications, communications signal processing, estimation and detection techniques, signal processing algorithms for flexible radios, cognitive radio, full-duplex radio, radio localization, and 5G mobile radio networks.
\end{IEEEbiography}




\end{document}